\documentclass[twocolumn,trackchanges,twocolappendix]{aastex701}

\usepackage{ragged2e}
\usepackage{amsmath}
\usepackage[nameinlink]{cleveref}
\usepackage{threeparttable}

\usepackage[]{natbib}
\setcitestyle{comma}
\usepackage{comment}
\usepackage{wasysym}

\usepackage[utf8]{inputenc}
\usepackage[T1]{fontenc}
\usepackage{lineno}[mathlines]

\makeatletter
\newcommand*{\linktocite}[2]{%
  \hyper@natlinkstart{#1}#2\hyper@natlinkend}
\makeatother

\def\K{\; {\rm K}}
\newcolumntype{P}[1]{>{\centering\arraybackslash}p{#1}}
\usepackage{siunitx}

\graphicspath{{./}{figs/}}

\begin{document}

\title{JWST MIRI reveals a potential atmosphere on the ultra-hot rocky planet TOI-431b}

\author[0009-0004-1094-3093]{Cole Smith}
\affiliation{Department of Astronomy, University of Maryland, College Park, MD 20742, USA}
\email{}

\author[0000-0002-0508-857X]{Brandon Park Coy}
\affiliation{Department of the Geophysical Sciences, University of Chicago, Chicago, IL 60637, USA}
\email{}
\author[0000-0003-4241-7413]{Megan Weiner Mansfield}
\affiliation{Department of Astronomy, University of Maryland, College Park, MD 20742, USA}
\email{}

\author[0000-0002-4671-2957]{Rafael Luque}
\affiliation{Instituto de Astrof\'isica de Andaluc\'ia (IAA-CSIC), Glorieta de la Astronom\'ia s/n, Granada, Spain}
\email{}
\author[0000-0002-4487-5533]{Anjali A. A. Piette}
\affiliation{University of Birmingham, School of Physics \& Astronomy, Birmingham, B15 2TT, UK}
\email{}
\author[0000-0002-6215-5425]{Qiao Xue}
\affiliation{Department of Astronomy \& Astrophysics, University of Chicago, Chicago, IL 60637, USA}
\email{}
\author[0000-0002-0659-1783]{Michael Zhang}
\affiliation{Department of Astronomy \& Astrophysics, University of Chicago, Chicago, IL 60637, USA}
\email{}
\author[0000-0003-4733-6532]{Jacob L.\ Bean}
\affiliation{Department of Astronomy \& Astrophysics, University of Chicago, Chicago, IL 60637, USA}
\email{}
\author[0000-0003-2404-2427]{Madison Brady}
\affiliation{Department of Astronomy \& Astrophysics, University of Chicago, Chicago, IL 60637, USA}
\email{}
\author[0000-0002-8518-9601]{Peter Gao}
\affiliation{Carnegie Science, Washington, DC 20015 USA}
\email{}
\author[[0000-0003-2775-653X]{Jegug Ih}
\affiliation{Space Telescope Science Institute, 3700 San Martin Drive, Baltimore, 21218, MD, USA}
\email{}
\author[0000-0002-1337-9051]{Eliza M.-R. Kempton}
\affiliation{Department of Astronomy \& Astrophysics, University of Chicago, Chicago, IL 60637, USA}
\email{}
\author[0000-0002-1426-1186]{Edwin Kite}
\affiliation{Department of the Geophysical Sciences, University of Chicago, Chicago, IL 60637, USA}
\email{}
\author[0000-0002-9076-6901]{Daniel Koll}
\affiliation{Peking University, Beijing, People's Republic of China}
\email{}
\author[0000-0003-2066-8959]{Jaume Orell}
\affiliation{Department of Astronomy, University of Texas at Austin, Austin, TX, USA}
\email{}
\author[0000-0003-0987-1593]{Enric Palle}
\affiliation{Instituto de Astrofísica de Canarias: San Cristobal de La Laguna, Canarias, ES}
\affiliation{Departamento de Astrof\'isica, Universidad de La Laguna (ULL), E-38206 La Laguna, Tenerife, Spain}
\email{}

\begin{abstract}

An open question in exoplanet science is whether rocky exoplanets extremely close in to their host stars, `lava worlds', can retain significant atmospheres. Thermal emission observations via secondary eclipse can be used to determine whether a rocky exoplanet possesses an atmosphere. It is expected that an atmosphere could increase the planet's albedo and/or redistribute heat away from the dayside, reducing the secondary eclipse depth measured. Recent eclipse observations of lava planets, rocky planets hot enough to have liquid magma surfaces, have suggested the presence of atmospheres. Here, we present a single JWST MIRI/LRS partial secondary eclipse of the lava planet TOI-431b. We measure an eclipse depth of 81 $\pm$ 16 ppm, which corresponds to a brightness temperature of $1967^{+237}_{-253}$ K and a brightness temperature ratio $\mathcal{R}$ = $0.81\pm0.10$, being 1.2 $\sigma$ higher than the value reported by Spitzer. The observed brightness temperature ratio is 1.8$\sigma$ below that of a zero-albedo, zero heat redistribution bare rock ($\mathcal{R}$=1). Given that magma pools are expected to have low albedos, we find that our results are best explained by the presence of an atmosphere. Future work  to measure the exact composition of TOI-431b's atmosphere and improve models of lava planet atmospheres would better constrain the nature and evolution of lava planets.

\end{abstract}


\section{Introduction}

The search for atmospheres on rocky exoplanets is of immense interest in helping to better characterize the population and their potential for habitability. Many rocky exoplanets, especially those that are ultra-hot or in close orbits, are thought to lose their primary atmospheres and may subsequently experience the removal of secondary atmospheres through stellar-driven photoevaporation \citep{Dong2018, Kite&Barnnett2020}. However, planets with an irradiation temperature above 1700 K (lava planets) may be hot enough to have their surfaces begin to evaporate and form a silicate vapor atmosphere in vapor pressure equilibrium with the surface \citep{Schaefer2009, Miguel2011}. The composition of these atmospheres remains poorly constrained, with rock vapor-rich and volatile-rich atmospheres allowed by various models depending on the interior composition and outgassing history \citep{Piette2023, Zilinskas2023}. Depending on the atmospheric composition, many of these volatiles could be potentially observable given enough signal to noise ratio. By studying these planets in secondary eclipse, we can begin to explore both the planet's atmosphere as well as its internal composition.

Secondary eclipse photometry can be an effective method for discerning whether these rocky planets host atmospheres \citep{Koll2019, Mansfield2019}. A thick atmosphere could impose heat redistribution away from the dayside of the planet as well as an elevated albedo from the presence of clouds \citep{Koll2016}. The planet's surface must also be taken into consideration. Lab measurements suggest that lava planets would have molten surfaces with low albedos and negligible heat redistribution \citep{Essack2020}. This means that large reductions in the day–night temperature contrast would likely only result from the presence of an atmosphere. An added benefit of secondary eclipse photometry is that it's not as prone to stellar contamination as transits \citep{Rackham2018}.

Recent secondary eclipse observations have suggested the presence of thick and/or reflective atmospheres on several lava planets. Observations of HD~3167b \citep{Coy2026}, TOI-561b \citep{Teske2025}, K2-141b \citep{Zieba2022}, and 55 Cancri e \citep{Hu2024, Patel2024}, planets with irradiation temperatures >2400\,K, have found low dayside temperatures indicating that these planets could potentially possess atmospheres. Observations of a larger population of lava planets would provide further insights into our understanding of their atmosphere.


TOI-431b \citep{Osborn2021} is an ultra-short period super-Earth ($R=1.3\,R_{\oplus}, P=0.49$\,d, $M = 3.07\,M_{\oplus}$) with an equilibrium temperature of 1862 K.  Notably, its density ($8\pm1$\,g/cm$^3$) is consistent with that expected of an Earth-like interior composition \citep{zeng2019} contrary to the slightly under-dense super Earths HD~3167b, 55~Cancri~e, and TOI-561b, making it a bonafide `lava world'. It orbits a K-type star and has two other confirmed planets as part of the system. Planet c is a non-transiting super-Earth ($P=4.85$\,d, $M\sin(i) = 2.83\,M_{\oplus}$), and Planet d is a transiting sub-Neptune ($R=3.29\,R_{\oplus}, P=12.46$\,d, $M = 9.9\,M_{\oplus}$). Previous observations with Spitzer IRAC Channel 2 at 4.5~$\mu$m observed an eclipse depth of 33 $\pm$ 22 ppm and a low brightness temperature of $1520^{+360}_{-390}$K \citep{Monaghan2025}, indicating the possible presence of an atmosphere. However, the detection of the eclipse was only made at 1.5$\,\sigma$ confidence after averaging seven separate eclipses.

In this paper, we present a JWST observation of TOI-431b in secondary eclipse thermal emission to search for a potential atmosphere to strengthen the tentative Spitzer detection. Section \ref{sec:observations} presents our observation details as well as our data reduction and modeling setup.  Section \ref{sec:Results} presents our analysis and results for the eclipse depth and extracted dayside surface temperature. The implications of our results and discussion of planned future work are also described in Section \ref{sec: Disc}. Finally, we summarize key points in Section \ref{sec:conc}.

\section{Observations \& Data Reduction}
\label{sec:observations}

We observed one secondary eclipse of TOI-431b with JWST MIRI/LRS on 2025 February 1 as part of the LAVA LAMPS survey (program GO 4818, M. Weiner Mansfield PI), with a total exposure duration of 3.75 hours. The observations cover a wavelength range of 5–12 $\mu$m using the Low Resolution Spectrometer (LRS) slitless time-series mode with subarray SLITLESSPRISM and the FASTR1 readout pattern. The observation used 8 groups/integration with a maximum saturation fraction of 74\%. The observation is a partial eclipse with the eclipse running into the end of the observation window. This occurred due to uncertainties in the ephemeris timing based on radial velocity measurements taken from \cite{Osborn2021} and results in no post-eclipse baseline. However, based on the eclipse time derived below, we observed 86\% of the eclipse.

\begin{figure*}[t!]
    \includegraphics[width=1\textwidth]{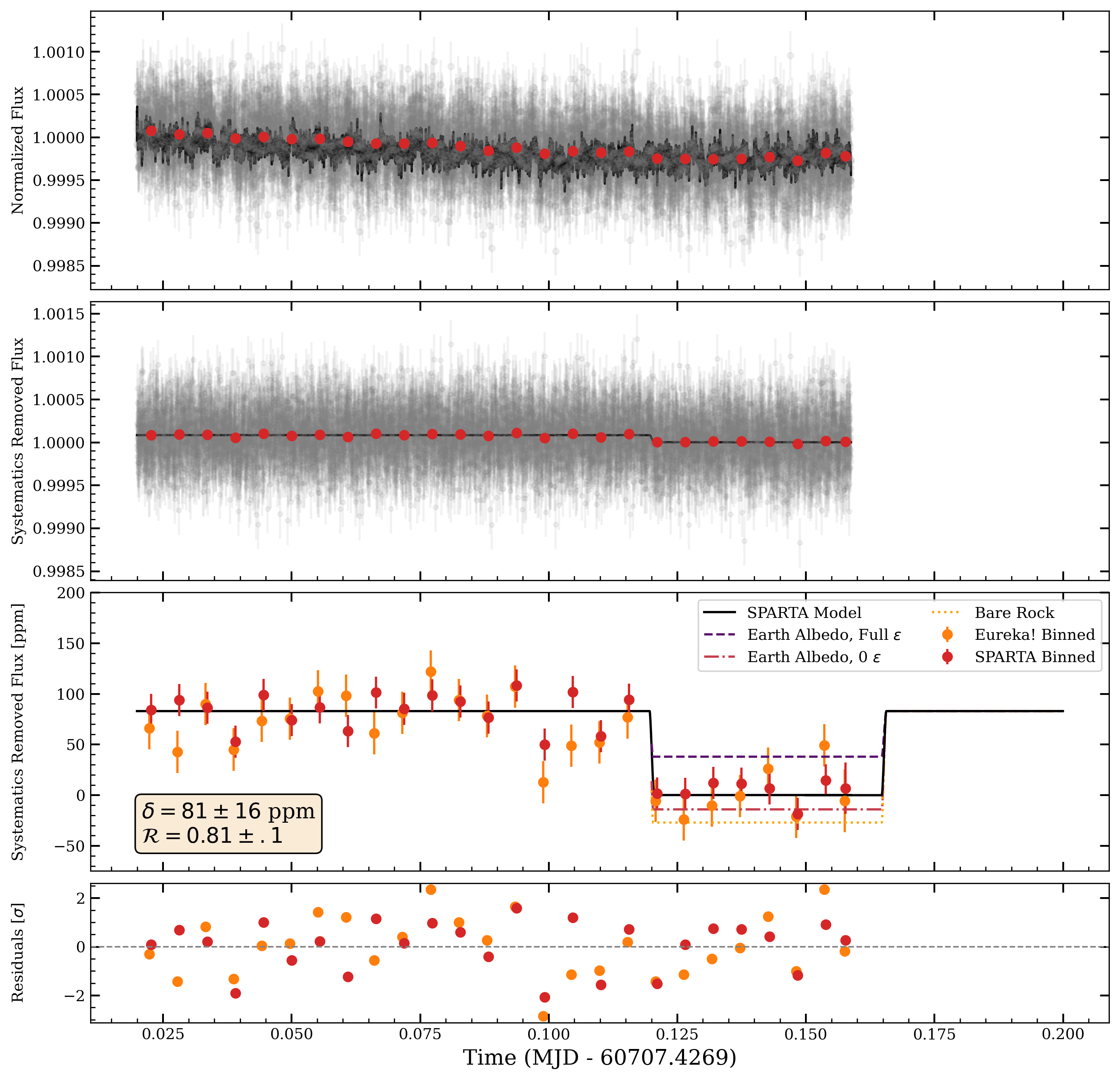}
    \centering
    \caption{The observed lightcurve for TOI-431b. \textit{First Panel:} The normalized flux \texttt{SPARTA} lightcurve with the binned points overlaid in red with respect to time. The determined eclipse and systematics model is overplotted in black. \textit{Second Panel:} The light curve with the systematics removed. \textit{Third Panel:} A zoom in on the binned points overlaying multiple light curves with varying eclipse depths, corresponding to different albedos and/or heat redistribution factors. These eclipse depths are derived from Equations \ref{eq:EclipseDepth} and \ref{eq:R_value} for the respective temperatures. Both the \texttt{Eureka!} and \texttt{SPARTA!} fits are plotted with the black model belonging to the \texttt{SPARTA} reduction. The models are extended beyond the end of our observation to show the full extent of the eclipse. \textit{Bottom Panel:} The residuals for the binned points with respect to each eclipse model in terms of $\sigma$.}
   \label{fig:LC}
\end{figure*}

To better constrain the mid-eclipse time and ensure that subsequent fitting to the light curve is accurate since there is no post-eclipse baseline, we first jointly modeled the TOI-431 system parameters with \texttt{juliet} \citep{Espinoza2019}, as described in Section \ref{subsec:juliet}. We then performed two independent data reductions of the JWST data, which are described in Sections \ref{subsec:SPARTA} and \ref{subsec:Eureka!}, and both reductions can be seen in Figure \ref{fig:LC}.

\subsection{Refining the system parameters}
\label{subsec:juliet}

Precise planet parameters are key in interpreting the eclipse depths of small planets, as the uncertainties in the derived dayside brightness temperature can often be dominated by orbital/planet parameter uncertainty \citep{coy2025population,monaghan2026uniform}.
\citet{Osborn2021} analyzed TESS photometry from Sectors 5 and 6 (Nov \& Dec 2018) and radial velocity measurements (HARPS, HIRES, iSHELL, FEROS, and Minerva-Australis). Their joint fit included only the highest-quality datasets, namely TESS, HARPS, and HIRES. In this work, taking advantage of additional TESS observations of the TOI-431 system during Sectors 32 (Dec 2020) and 98 (Dec 2025), we reproduce the joint fit in \citet{Osborn2021} using \texttt{juliet} to update key planetary parameters. 

Our setup is nearly identical to that of \citet{Osborn2021}. Following their analysis, we include only TESS (with the addition of Sectors 32 and 98) photometry and radial velocities from HARPS and HIRES. We include three Keplerian orbits with a prior on the orbital eccentricity $e$ following a Beta distribution with $\alpha=1.52$ and $\beta=29$ for the two outer planets (c and d), as expected for tightly-packed multi-transiting systems \citep{vanEylen2019}. A Beta distribution is a continuous probability distribution defined to be between 0 and 1 and is shaped to favor smaller values of orbital eccentricity $e$ if $\beta > \alpha$ and vice versa. We experimented fits with eccentricity as free parameters for planet b as well but ultimately determined that the eccentricity derived from our \texttt{juliet} fit and secondary eclipse timing with JWST are fully consistent with 0 and thus fix $e_{b}=0$. We take advantage of the multi-planet nature of the TOI-431 system by fitting for the stellar density $\rho_{\star}$ instead of the individual planets' semi-major axes ($a/R_{\star}$), which helps break degeneracies between $a/R_{\star}$ and orbital inclination \citep{seager2003unique}.  We use a wide Gaussian prior of $\rho_{\star}=2.81\pm0.55$ g/cm$^{3}$ based on the stellar mass and radius constraints reported in \citet{Osborn2021}.  We tested including an uninformative wide log-uniform prior for $\rho_{\star}$ and retrieve near identical but slightly less precise results. We adopted a quadratic limb darkening law for TESS data using the $q_1, q_2$ parametrization introduced by \citet{Kipping2013}.  We use Gaussian priors for $q_{1}$ and $q_{2}$ with widths of 0.1 centered on the expected values for TOI-431 output by \texttt{ExoTIC-LD} \citep{Grant2024} using the Stagger \citep{magic2015stagger} 3D grid of stellar atmosphere models.  To model stellar variability, we use Gaussian Processes (GPs) for both the photometric and radial velocity data, although they do not share any hyperparameters. For the photometry, we use \texttt{celerite}'s M\'atern 3/2 kernel of the form
\begin{equation*}
    k_\tau = \sigma^2_\mathrm{GP,TESS} \left( 1 + \frac{\sqrt3\tau}{\rho_\mathrm{GP,TESS}} \right)\exp\left(-\frac{\sqrt3\tau}{\rho_\mathrm{GP,TESS}} \right),
\end{equation*}
where $\tau = |t_{i} - t_{j}|$ is the time-lag, $\rho_\mathrm{GP,TESS}$ the characteristic timescale, and $\sigma_\mathrm{GP,TESS}$ the amplitude of the GP modulation. Contrary to \citet{Osborn2021}, we fit separate GP hyperparameters for each TESS sector (treating sectors 5 \& 6 as a continuous sector), as the photometric variability of TOI-431 changes drastically between sectors.  For the radial velocities, we use \texttt{celerite2}'s \citep{celerite2} quasi-periodic kernel of the form
\begin{equation*}
k_\tau = \frac{B}{2+C}e^{-\tau/L}\left[\cos \left(\frac{2\pi \tau}{P_\textnormal{rot}}\right) + (1+C)\right],
\end{equation*}
where $\tau = |t_{i} - t_{j}|$ is the time-lag, $B$ and $C$ define the amplitude of the GP, $L$ is a timescale for the amplitude-modulation of the GP, and $P_\textnormal{rot}$ is the rotational period of the modulations, on which we put a tight prior based on the stellar rotation determined by \citet{Osborn2021}. Finally, we added an instrumental jitter term for each instrument. 

The joint fit model parameters, priors, and posterior distributions are reported in Table~\ref{tab:posteriors}, while Table~\ref{table:params_derived} reports the derived system parameters. As expected, the results are similar to those reported in \citet{Osborn2021}, although we improve the precision on TOI-431b's orbital period by more than two orders of magnitude ($\sim6\times$ for TOI-431d) and report the (negligible) eccentricities of planets c and d for the first time, both of which are consistent with zero (99\% confidence intervals of $e_{c}<0.14$ and $e_{d}<0.10$).

\subsection{JWST data reduction with \texttt{SPARTA}}
\label{subsec:SPARTA}
\texttt{SPARTA} is an end-to-end JWST data reduction pipeline, first presented in \cite{Kempton2023}, that is completely independent of the standard \texttt{jwst} pipeline. Detailed documentation of the version used for this work can be found in \cite{xue2025jwst}. We begin by performing nonlinearity correction, dark subtraction, multiplication by the gain (assumed to be 3.1 electrons per data number, \citealt{Kempton2023}), two rounds of up-the-ramp fitting to take out MIRI LRS nonlinearities, and flat fielding.  \texttt{SPARTA} rotates the MIRI image so that the wavelength axis spans the $x$ direction.  We then remove the background at each wavelength column by subtracting the median of rows 10:25 and -25:-10 for each column.

We use these corrected images to calculate a median image across all integrations and use it as a template to calculate the position offset of the trace in each integration in both the $x$ and $y$ directions. Using the previously calculated median image to generate a spatial profile, we use optimal extraction \citep{horne1986optimal} to extract a spectrum for each integration. Following previous studies using \texttt{SPARTA} with MIRI LRS \citep{Xue2024,WeinerMansfield2024,Coy2026}, we use an extraction half-width of 5 pixels and reject pixels more than 5$\,\sigma$ away from the reconstructed image as outliers.

 Integrations that are $>4\,\sigma$ outliers in the white light curve are rejected, while $>4\,\sigma$ outliers in the unbinned spectroscopic light curves are repaired via interpolation.  For the white light curve, we sum the data over 5.06--10.55\,$\mu$m, motivated by the mismatch between model and observed stellar fluxes at wavelengths shorter than 5.06\,$\mu$m (e.g., \citealt{zieba2026dark}).  We also notice a sharp change in the ramp behavior near the `shadowed region' of the slitless prism (10.55-11.77\,$\mu$m, \citealt{bell2024nightside}).

To model the white light curve, we use nested sampling with \texttt{dynesty} \citep{speagle2020dynesty} and a convergence criterion of $\Delta$log$Z$=0.1.  We use dynamic nested sampling with 1000 initial live points, as it is designed to better-describe the true posterior distribution than static nested sampling.

We fix $a/R_{\star}$, $R_{p}/R_{\star}$, and the orbital inclination to the median parameters from our \texttt{juliet} fit (Table \ref{tab:posteriors}), and assume an orbital eccentricity of zero.  We also include correct for the light travel time delay (roughly a maximum of $2a/c=11$ seconds) by fixing the stellar radius to 0.731$\,R_{\odot}$ \citep{Osborn2021}.  We tested allowing for a free mid-eclipse time and imposing a Gaussian prior based on the propagated uncertainty on ephemeris from our \texttt{juliet} fit ($\sim$0.7 minutes), but ultimately fixed the mid-eclipse time (to BMJD 60707.56952666) due to the lack of post-eclipse baseline.  Allowing for a free mid-eclipse time using a wide uniform prior fits a mid-eclipse time $-0.1^{+0.4}_{-1.5}$ minutes from our imposed timing, highlighting that we are confidently detecting the eclipse despite the lack of post-eclipse baseline.

Long-term systematic trends in the white light curve are modeled via:
\begin{equation}
    F_{sys}(t)=F_{star}\times(1+A(t-t_{min})^2+B(t-t_{min})+c_{x}x),
\end{equation}
where $t_{min}$ represents the beginning time of the trimmed light curve, $F_{star}$ is a normalization constant, and $c_{x}$ decorrelates with the $x$ position of the spectral trace (the spectral direction; referred to as $y$ in \texttt{Eureka!}). We do not find significant correlation with the $y$ position of the spectral trace and thus do not include it in our fits.  We trimmed the first 1000 integrations (24 minutes) of the white light curve, determined by maximizing the normalized log-likelihood as described in \citet{Coy2026}. We tested including a quadratic (above), linear, and linear+exponential systematics model by comparing the Bayesian information criteria (BIC) of each best-fit model.  The quadratic model is by far the most favored, with a $\Delta$BIC of 902 and 21 compared to the linear+exponential and linear models, respectively.  The linear+exponential model finds a similar eclipse depth of $77\pm15$ ppm, but fits ramp timescales much longer than the duration of the observation ($\tau\sim1.3$ days).

Our astrophysical model is determined using \texttt{batman} 
\citep{Kreidberg_2015}.  We tested including a phase modulation term due to the planet's short orbital period, where
\begin{equation}
    F_{p}(\theta)=F_{p,batman}\times\left(\frac{1}{2}+\frac{1}{2}\cos{\theta}\right),
\end{equation}
where $\theta$ is the planet's orbital phase relative to mid-eclipse.  We retrieve an eclipse depth of $81\pm16$ ppm both with and without including phase modulation. This is likely due to the planet's actual phase modulation being absorbed into the quadratic ramp systematics model, and thus we exclude this term in our final fits.  Our eclipse depth precision is noticeably worse than the photon-limited precision expected from JWST ETC output ($\sim11$\,ppm), most likely due to our lack of post-eclipse baseline.  We investigated the RMS versus bin-size of the best-fit light curve residuals and found no evidence for additional correlated noise.

We extracted spectral light curves using the same systematics model described above for five bins (5.05-6.15, 6.15-7.25, 7.25-8.35, 8.35-9.45, and 9.45-10.55 $\mu$m respectively), but the mid-eclipse time was fixed to the best-fit mid-eclipse time from the white light curve.

\subsection{JWST data reduction with \texttt{Eureka!}}
\label{subsec:Eureka!}

\texttt{Eureka!} is a 6-stage data-reduction pipeline for analyzing exoplanet transits and secondary eclipses, described in detail in \cite{Bell:2022}. The first two stages follow those in the \texttt{jwst} pipeline, and we used a $\texttt{jump\_detection}$ = 4.0 parameter for Stage 1 as varying the value did not have significant impacts on the output white light curve. Stage 3 handles background subtraction and optimal spectral extraction. Similarly to the \texttt{SPARTA} reduction, we extract the white light curve over a wavelength range of  5.05 to 10.55 \,$\mu$m to avoid the shadowed region. Optimal spectral extraction \citep{horne1986optimal} was performed with a half-width of 6 pixels around the trace, and background subtraction was performed using pixels >=15 pixels away from the trace. This aperture size was chosen as it minimized the MAD, but other aperture sizes also produced eclipse depths that were consistent within  1$\sigma$. We used a 5$\sigma$ threshold for outlier rejection during optimal spectral extraction. Along the time axis, an outlier rejection routine was applied with a double-iteration sigma threshold [5,5], which means any points that strayed beyond 5$\,\sigma$ from the mean were not considered.

Stage 4 of the \texttt{Eureka!} pipeline separates the data into spectroscopic light curves, and Stage 5 performs light curve fitting. For Stage 5, we fix all stellar parameters to their literature values in \cite{Osborn2021} and use the \texttt{juliet} derived median values for $a/R_{\star}$, $R_{p}/R_{\star}$, and inclination with a fixed eccentricity of 0. The secondary eclipse timing was tested using two approaches. The first was a fully unconstrained approach that failed to find the eclipse. The second was to use the \texttt{juliet} derived mid-eclipse timing with a Gaussian prior. This found the eclipse and the derived timing was consistent with the \texttt{juliet} timing within 1$\sigma$. We fit instrument systematics with a temporal quadratic model (three free parameters) and removed the first 24 minutes (1000 integrations) of observation due to extreme ramping. We performed a MCMC fit using \texttt{emcee} \citep{Foreman-Mackey2013}. We ran 200 walkers for 25,000 steps each, discarding the first 500 as burn-in. The mean integrated autocorrelation time across all parameters was 54 steps, giving an effective sample size of ~45,000, and the chains exhibited a healthy mean acceptance fraction of 0.55. These diagnostics indicate that the MCMC chains were well mixed and fully converged.

\begin{figure*}[t!]
    \includegraphics[width=.87\textwidth]{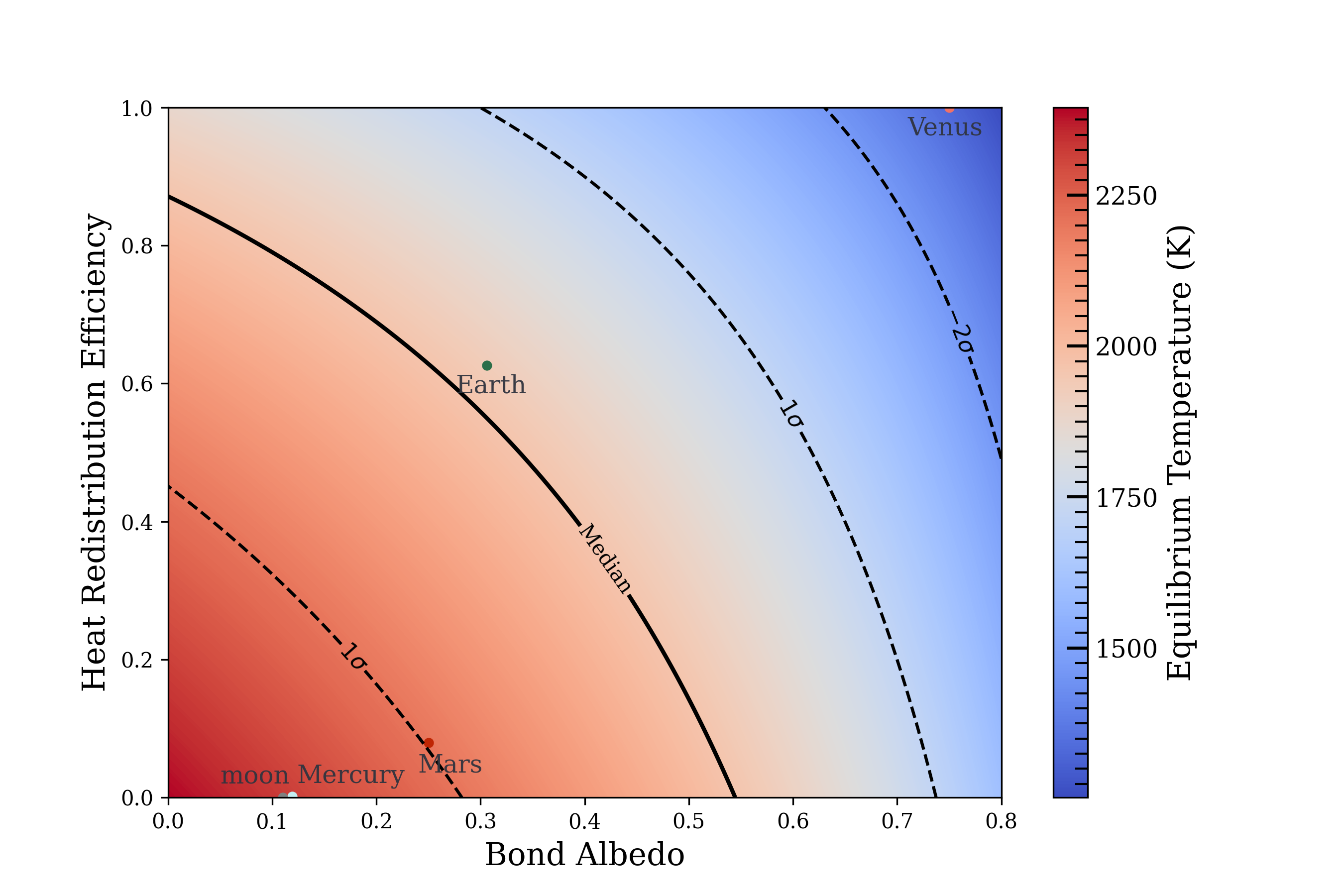}
    \centering
    \caption{Contour plot showing the brightness temperature of TOI-431b in the LRS bandpass as a function of the planet’s Bond albedo and heat redistribution efficiency. The solid black line is the derived dayside temperature in the LRS bandpass, and the dotted lines show $\pm$1 and 2$\sigma$ dayside temperatures respectively. This shows that the planet is more than 1.8$\sigma$ inconsistent with being a bare dark rock. Colored points show Bond albedo and heat redistribution values for solar system bodies, adopted from \cite{Xue2024}.}
   \label{fig:Albedo_vs_f}
\end{figure*}

The measured eclipse depth using \texttt{Eureka!} was 73 $\pm$ 21 ppm. To determine the sensitivity of our eclipse depth to the chosen systematics model, we also tested a purely linear fit as well as linear fit with an initial exponential ramp. The quadratic model fit was preferred with a $\Delta$BIC of 15 and 21 respectively. The exponential ramp and quadratic fits produced eclipse depths that were consistent within $1\sigma$. As an additional test, we modified how many integrations from the beginning of the observation were trimmed and found eclipse depths consistent within 1$\sigma$.

\section{Analysis \& Results}
\label{sec:Results}

The two data reductions agree well within 1$\sigma$, so from this point on all calculations use the \texttt{SPARTA} reduction, which had a slightly lower white light eclipse depth error.

\subsection{Brightness Temperature Ratio}

We calculate the dayside brightness temperature of TOI-431b via the equations presented in \cite{Xue2024}:

\begin{equation}
\label{eq:EclipseDepth}
\frac{F_p}{F_\star} = \left(\frac{R_p}{R_\star}\right)^2 \cdot \frac{\int \frac{\pi B_p(T_p, \lambda)}{hc/\lambda} \cdot W_{\lambda} d \lambda}{\int \frac{M_\star(T_\star, log \ g,[M/H], \lambda)}{hc/\lambda} W_{\lambda} d \lambda}
\end{equation}
\noindent where $W_\lambda$ is the throughput of MIRI/LRS, $B_p(T_p, \lambda)$ is the blackbody Planck function, $M_\star(T_\star, log \ g,[M/H], \lambda)$ is the stellar spectrum, $T_\star$ is the stellar temperature, $log \ g$ is the stellar surface gravity, and $[M/H]$ is the stellar metallicity. For the stellar spectrum, we used models interpolated from the PHOENIX grid \citep{Allard2012} using the Python package \texttt{pysynphot15} \citep{pysynphot2013}.

From the dayside brightness temperature, we calculate the Bond albedo and heat recirculation efficiency \citep{Cowan&Agol2011}:

\begin{equation}
\label{eq:T_irr}
T_{irr} = T_\star\sqrt{\frac{R_\star}{a}}
\end{equation}

\begin{equation}
\label{eq:T_Dayside}
T_{p, dayside} = T_{max}\  \cdot \mathcal{R} = \left(\frac{2}{3}\right)^{\frac{1}{4}} \ \cdot T_\star\sqrt{\frac{R_\star}{a}}\  \cdot \mathcal{R}
\end{equation}

\begin{equation}
\label{eq:R_value}
\mathcal{R} = \left(\frac{2}{3}\right)^{-\frac{1}{4}} \ \cdot (1 - A_B)^{\frac{1}{4}} \ \cdot \left(\frac{2}{3} - \frac{5}{12} \epsilon\right)^{\frac{1}{4}} ,
\end{equation}

\noindent where $T_{irr}$ is the irradiation temperature, $R_\star$ is the radius of the host star, $T_\star$ is the temperature of the host star, $T_{p, dayside}$ is the extracted dayside temperature of the planet, $A_B$ is the Bond albedo of the planet, and $\epsilon$ is the heat redistribution efficiency, defined as being 1 for full planetary redistribution and 0 for no redistribution from dayside to nightside. These equations are normalized so that $\mathcal{R}$ = 1 for a planet with zero albedo and zero heat redistribution.

Using Equation \ref{eq:EclipseDepth}, we measured a dayside temperature of $1967^{+237}_{-253}$K.  The MIRI LRS white light (5.06--10.55\,$\mu$m) eclipse depth for a maximally hot dayside ($\mathcal{R}=1$) on TOI-431b would be $112_{-4}^{+5}$\,ppm, corresponding to a dayside temperature of $2400\pm50$ K. Our measured eclipse depth corresponds to a brightness temperature ratio of $\mathcal{R}=0.81\pm0.10$. Using this value, we rule out a zero albedo, zero heat redistribution dayside at a confidence of $1.8\sigma$. 

\cite{Monaghan2025} observed an eclipse depth of 33 $\pm$ 22 ppm for TOI-431b using Spitzer (4.5 $\mu\,$m) with a corresponding brightness temperature of $1520^{+360}_{-390}$K, which falls within 1.2$\sigma$ of our observation. The improved precision on the brightness temperature measured with JWST compared to Spitzer increases our confidence that TOI-431b has a dayside temperature lower than the zero-albedo, zero-redistribution maximum.

Figure \ref{fig:Albedo_vs_f} shows the Bond albedos and heat redistribution factors consistent with the band-integrated eclipse depth we observe. TOI-431b's temperature ratio $\mathcal{R}$ is most consistent with the expected brightness temperature ratio of an Earth-like atmosphere. However, given that TOI-431b's temperature far exceeds that of any body in the solar system, we explored in detail a broad sample of potential lava surfaces and atmosphere conditions to determine what best matched the observed dayside temperature.

\subsection{Lava surface models}

One potential explanation for the reduced eclipse depth of TOI-431b is the presence of a high albedo surface. If we assume there is zero heat redistribution on the planet, an albedo of $\sim$0.55 would be necessary to explain our observed dayside temperature. Given the high surface temperature of TOI-431b, it is likely its surface is at least partially melted. Liquid lava on the surface of the planet is expected to be dark and exhibit a relatively low albedo \citep{Essack2020}. For a completely molten surface, the geometric albedo is expected to reach at most $\sim$0.1 \citep{Essack2020}, corresponding to a maximum Bond albedo of $\sim$0.15 under the assumption of Lambertian scattering. As temperatures rise and melting becomes more widespread, both reduced reflectivity and a plateau in emissivity are expected \citep{Petrov2007, Dvurechensky1979}. As such, we believe that it is unlikely that the surface is reflective enough to reproduce the necessary albedo for these observations.

\subsection{Atmospheric Models}
\label{subsec:Atmos}

\begin{figure*}[t!]
    \includegraphics[width=.98\textwidth]{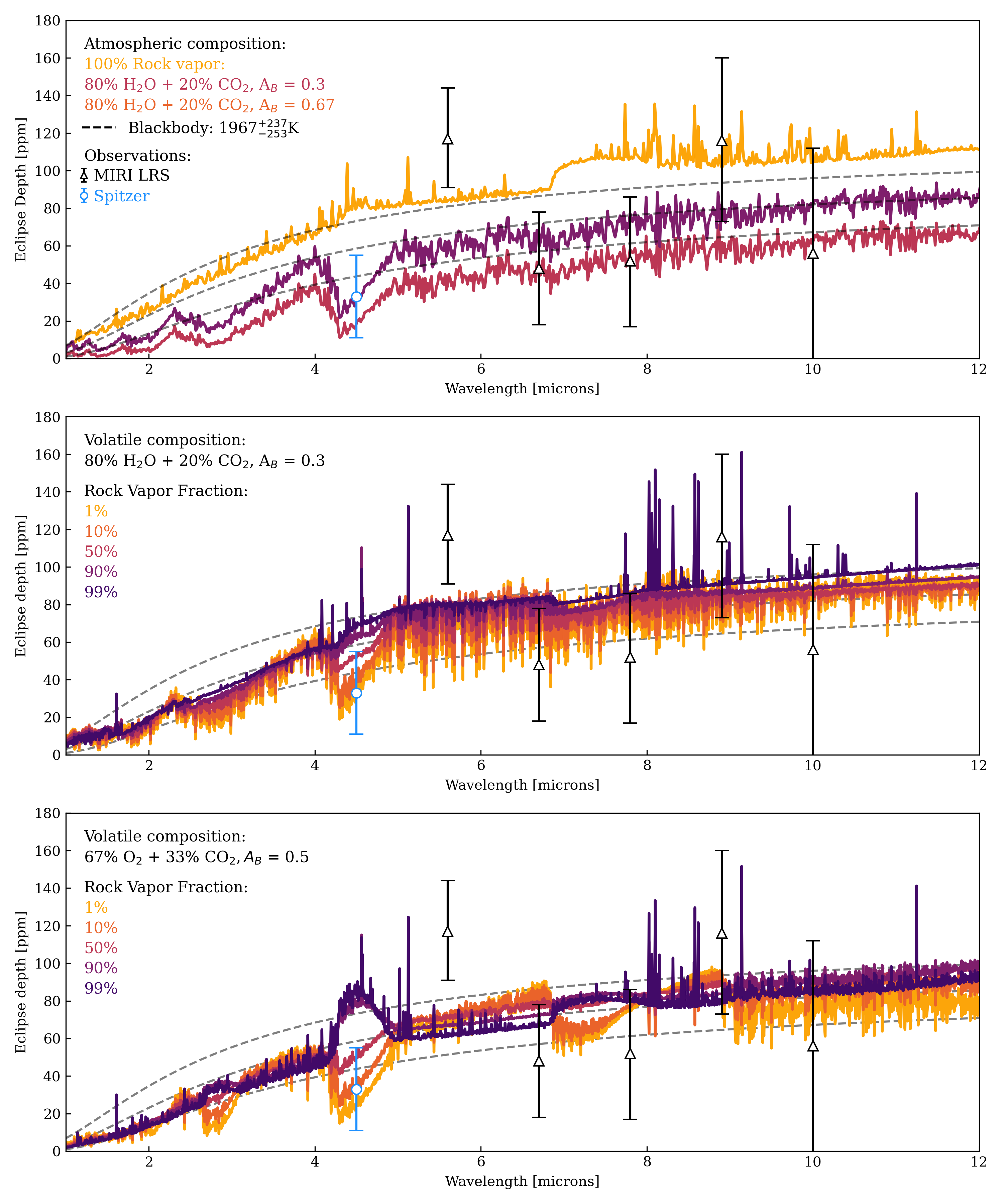}
    \centering
    \caption{Observed emission spectrum of TOI-431b from this work binned to 5 points in wavelength (5.05-6.15, 6.15-7.25, 7.25-8.35, 8.35-9.45, and 9.45-10.55 $\mu$m respectively) plotted in black triangles. The eclipse depth measured with Spitzer in \cite{Monaghan2025} is shown as a blue circle. Solid, colored lines show \texttt{GENESIS} models \citep{Piette2023} described in Section \ref{subsec:Atmos}. The dotted black lines shows the blackbody temperature derived from the white light eclipse depth with a $\pm$1$\sigma$ range. Multiple different atmospheres can produce an eclipse depth consistent with our observations, and future observations will be necessary to distinguish them.}
   \label{fig:ObsNeeded}
\end{figure*}

Given that a lava surface does not provide a good match to our observations, we examined what the white light eclipse depth and planetary spectrum would be for varying atmospheric compositions. Recent work has suggested that lava planets may have silicate-rich or volatile-rich atmospheres \citep{Zilinskas2022, Zilinskas2023, Maurice2024, Piette2023}, so we explored a variety of atmospheric compositions to determine both what could be consistent with our measured dayside temperature as well as what observations would be needed to discern between them. 

We model silicate-rich and volatile-rich atmospheres with \texttt{GENESIS} \citep{Gandhi2017,Piette2020_GENESIS} following the methods in \cite{Piette2023}. \texttt{GENESIS} solves for radiative-convective, hydrostatic, local thermodynamic, and thermochemical equilibrium to determine the temperature profile and emergent spectrum of the model atmosphere. It is coupled with the \texttt{FastChem} chemical equilibrium code \citep{Stock2022_fastchem} and the \texttt{VapoRock} magma ocean outgassing code \citep{Wolf2022_vaporock}.

In the \texttt{GENESIS} atmospheric we include opacity from key molecular, atomic and ionic species from the UV to the infrared. We calculate absorption cross sections for the molecular species using the methods of \citet{Gandhi2017} and using line list data from the ExoMol, HITEMP and HITRAN databases (details below). We consider opacity from the following molecular species and corresponding line lists: H$_2$O \citep{Rothman2010}, CO$_2$ \citep{Rothman2010}, CO \citep{Rothman2010}, SiO \citep{Yurchenko2021}, SiO$_2$ \citep{Owens2020}, AlO \citep{Patrascu2015}, MgO \citep{Li2019}, NaO \citep{Mitev2022}, TiO \citep{McKemmish2019}, O$_2$ \citep{Gordon2017}, OH \citep{Rothman2010}, FeH \citep{Dulick2003,Bernath2020}, NaH \citep{Rivlin2015}, NaOH \citep{Owens2021} and KOH \citep{Owens2021}. For the atomic and ionic species, we use opacities from the \texttt{DACE}\footnote{\url{https://dace.unige.ch/}} database, calculated using \texttt{helios-k} \citep{Grimm2021} and data from the Kurucz\footnote{\url{http://kurucz.harvard.edu/}} database \citep{Kurucz2018}. We consider opacity from several atomic and ionic species: Al, Ca, Fe, H, K, Mg, Na, O, Si, Ti, Ca$^+$, and Na$^+$. When rock vapor is included, it consists of a melt pool composition of the Bulk Silicate Earth, i.e., 45.97\% SiO$_2$, 36.66\% MgO, 8.24\% FeO, 4.77\% Al$_2$O$_3$, 3.78\% CaO, 0.35\% Na$_2$O, 0.18\% TiO$_2$, and 0.04\% K$_2$O \citep{Schaefer2009}.

Recent theoretical studies have explored the evolution of lava planets and disagreed on the predicted compositions of their atmospheres. For example, \cite{Curry2025} uses a simple chemical model of the dayside lava pool atmosphere system under mass-loss. By investigating the melting of the surface and its interaction with both the atmosphere as well as the mantle of the planet, they find that, in the regime of extreme mass loss, the system reaches an equilibrium where the atmosphere is composed of roughly the same material as the mantle. This suggests that probing the atmospheric spectrum of the planet could potentially gain information about its internal composition. However, \cite{Kite2016} find that if dense residual liquid drains back into the planet's interior, then the atmospheric composition can diverge from the mantle composition. 

Given the varied predictions for the compositions of lava planet atmospheres, we explore a broad range of models to determine what is consistent with our observations. For all models, we modified the Bond albedo to ensure that the broadband eclipse depth is consistent with our observations. Bond albedo is imposed by decreasing the incident flux at the top of atmosphere. The albedo may be considered as a stand-in for cloud condensation and/or heat redistribution, both of which are not included in this model but would decrease the dayside temperature. For optically thick models, the emergent spectrum is insensitive to surface pressure \citep[as long as this is deeper than the photosphere,][]{Piette2023}. In these models, we therefore choose a nominal surface pressure of 10 bars. The first case was an atmosphere whose volatiles were comprised of 67\%\,O$_2$ + 33\%\,CO$_2$ with a Bond albedo of 0.5 and a rock vapor fraction varying from 1\% up to 99\% (O$_2$-rich). The second case was an atmosphere whose volatiles were composed of 80\% H$_2$O + 20\% CO$_2$ with a bond albedo of 0.3 and a rock vapor fraction varying from 1\% up to 99\% (H$_2$O-rich). The remaining three cases involved either having 100\% rock vapor, or 0\% rock vapor and a purely volatile atmosphere of 80\% H$_2$O + 20\% CO$_2$ with Bond albedos of 0.3 and 0.67 respectively. All of these atmospheres with the exception of the pure rock vapor atmosphere and the 0.67 Bond albedo,  entirely volatile case create temperatures that are consistent within 1 $\sigma$ of the measured brightness temperature. The spectrum for each atmosphere and the extracted spectrum from our observation are shown in Figure \ref{fig:ObsNeeded}. 

These simple models suggest that the atmosphere must contain some volatile fraction to match our observed white light eclipse depth. However, a pure silicate atmosphere could be possible if it hosted a high-albedo silicate cloud deck, which would increase the Bond albedo and decrease the observed eclipse depth. Clouds are not included in these models, as their overall impact and likelihood of occurrence are poorly constrained.  Comprehensive modeling efforts will therefore be necessary to determine the possible cloud formation mechanisms in this environment and their resulting effects on the observed flux. 

\begin{figure*}[t!]
    \includegraphics[width=.87\textwidth]{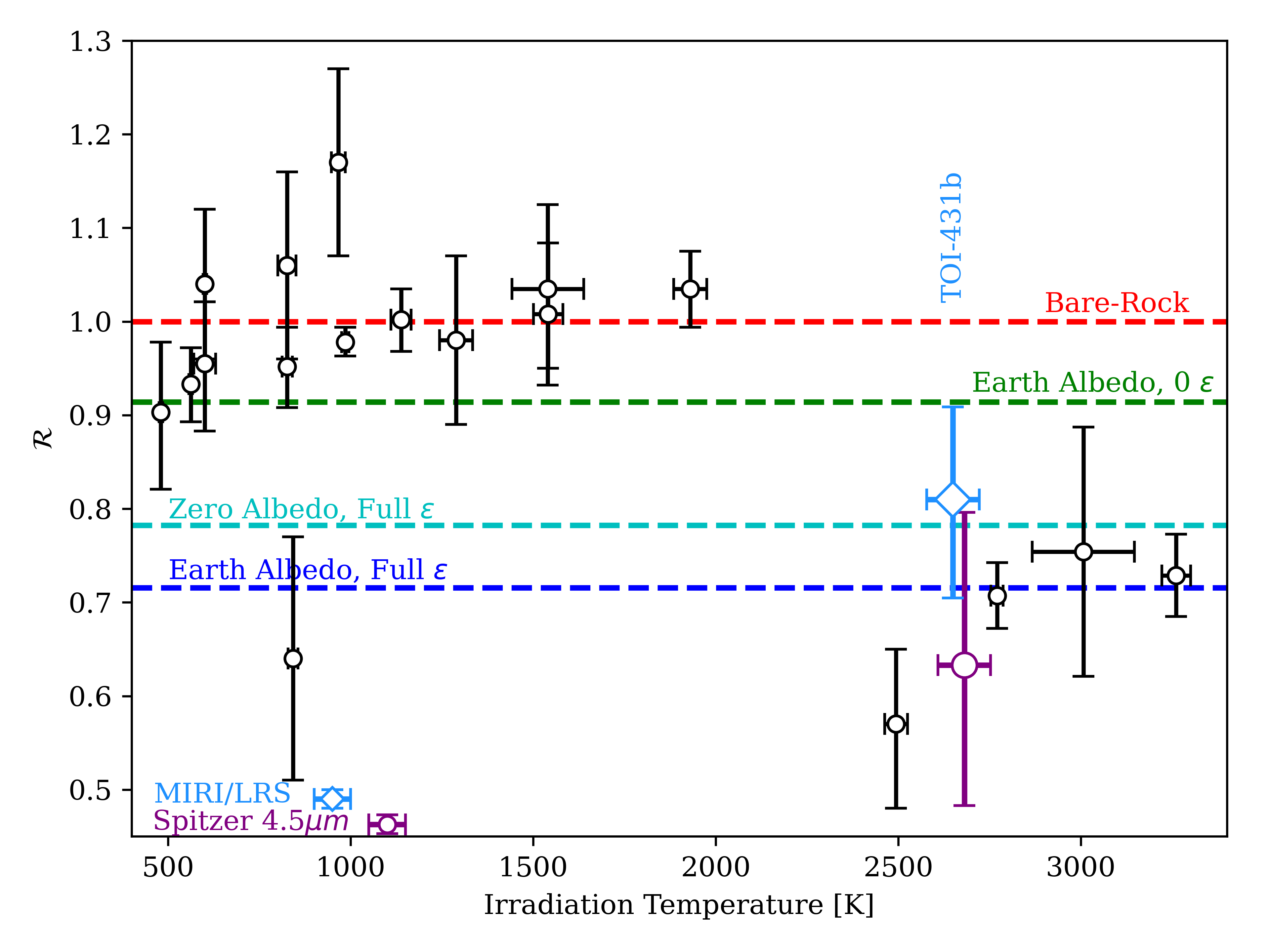}
    \centering
    \caption{Literature values for $\mathcal{R}$ (ratio of the dayside brightness temperature and the theoretical maximum) for all rocky exoplanets with observed dayside temperatures compared to TOI-431b. 
    Dashed horizontal lines show the respective $\mathcal{R}$ values for a bare, dark rock (red), an Earth albedo with zero heat redistribution (green), zero albedo with full heat redistribution (cyan), and an Earth albedo and full redistribution (dark blue). The $\mathcal{R}$ value derived within this work is plotted as a light blue diamond, and the value derived from \cite{Monaghan2025} using 4.5$\mu$m Spitzer observations is plotted as a purple circle and has a 30 K offset for visibility. TOI-431b matches a trend seen in other ultra-hot rocky exoplanets, with an $\mathcal{R}$ value that is noticeably lower than 1. The other planets plotted are from left to right: TRAPPIST-1c (using data from \citealt{Zieba2023, coy2025population}), TRAPPIST-1b \citep{Greene2023, Ducrot2025, coy2025population}, LTT 1445 A b \citep{Wachiraphan2025, coy2025population}, LHS 1140 c \citep{Fortune2025, Lin2026}, GJ 3929 b \citep{Xue2025, Lin2026}, GJ 1132 b \citep{Xue2024, coy2025population}, LHS 1478 b \citep{August2025}, TOI-1468 b \citep{MeierVald2025, Lin2026}, GJ 486 b \citep{WeinerMansfield2024, coy2025population}, LHS 3844 b \citep{Kriedberg2019, coy2025population}, LTT 3780 b \citep{Allen2025, Lin2026}, GJ 1252 b \citep{Crossfield2022, coy2025population}, TOI-1685 b \citep{Luque2025, coy2025population}, GJ 367 b \citep{zhang2024gj, coy2025population}, HD 3167 b \citep{Coy2026}, 55 Cnc e \citep{Hu2024}, K2-141 b \citep{Zieba2022}, and TOI-561b \citep{Teske2025}.}
   \label{fig:Population}
\end{figure*}

\begin{figure*}[t!]
    \includegraphics[width=\textwidth]{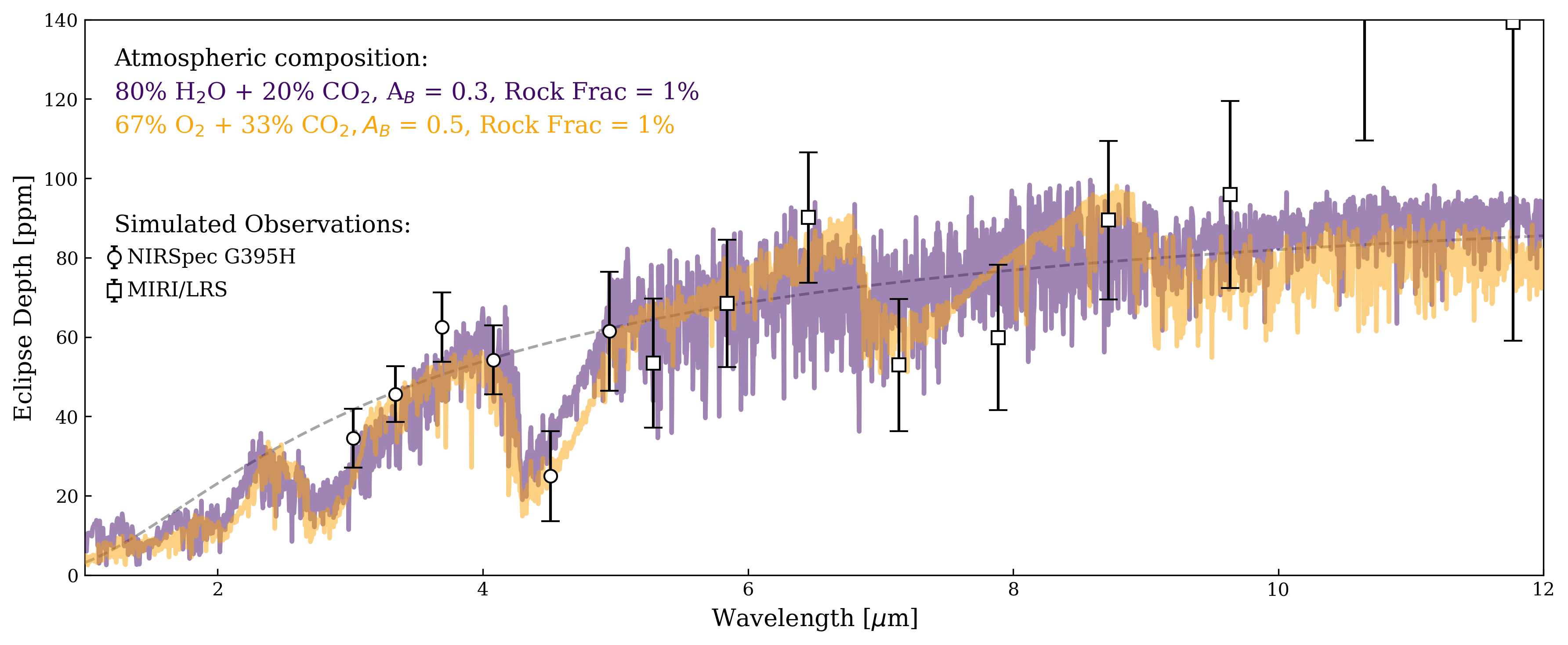}
    \centering
    \caption{Simulated GenTSO \citep{Cubillos2024} observations for TOI-431b compared to the H$_2$O (purple) and O$_2$-rich (yellow) \texttt{GENESIS} \citep{Piette2023} modeled atmospheres with 1\% rock vapor fractions. The simulated points plotted are for the O$_2$-rich atmosphere. A blackbody eclipse spectrum for the measured brightness temperature is overlaid in a dashed black line. The modeled observations are scaled and use 1 secondary eclipse observation following the procedures described in \ref{subsec:Sim_Obs}. Simulations modeled for NIRSpec G395H and MIRI/LRS are plotted in black circles and squares respectively.}
   \label{fig:Sim_Obs}
\end{figure*}

For cases including rock vapor, a clear SiO$_2$ emission feature is seen at around 7.5\,$\mu$m, and a CO$_2$ absorption feature is seen in the volatile rich cases near 4.3\,$\mu$m.  Future observations with JWST MIRI or NIRSpec could distinguish between these cases by probing these distinct features. For the H$_2$O-rich case, we can see that the presence of H$_2$O in the atmosphere completely washes out the silicate feature around 7.5\,$\mu$m. In the O$_2$-rich case, we see that the 7.5\,$\mu$m feature still persists, switching between being in either emission or absorption based on the rock vapor percentage. A similar trend is seen for the 4.3\,$\mu$m feature. By combining high SNR spectra in both MIRI and NIRSpec, the atmospheric composition could be more precisely determined.

\section{Discussion}
\label{sec: Disc}

\subsection{Lava Exoplanet Population}

TOI-431b joins a population of lava exoplanets with measured dayside temperatures that are below their expected maximum for a bare, dark rock, including HD~3167~b, K2-141~b, 55~Cancri~e, and TOI-561~b. Literature values for all known rocky exoplanets with reported dayside temperatures are plotted in Figure \ref{fig:Population} with our measurement for TOI-431b highlighted in blue. This trend suggests the presence of lower dayside temperatures on planets with higher irradiation temperatures, supporting the idea that atmospheres are common on lava planets. Putting TOI-431b in the context of these other ultra-hot rocky exoplanets increases our confidence in the measurement. 

\subsection{Future Work and Observations}
\label{subsec:Sim_Obs}

To motivate future observations and constrain possible atmospheric compositions, we simulated further observations of TOI-431b's spectrum. These observations were generated using GenTSO \citep{Cubillos2024}, a graphical interface that uses JWST’s Pandeia ETC engine to simulate observations for both MIRI/LRS and NIRSpec G395H. An observation duration of 4.76 hours was used with an in-eclipse duration of 1.012 hours. We used 8 and 3 groups per integration for MIRI and NIRSpec respectively. In both the H$_2$O and O$_2$-rich cases with rock vapor, the 1\% rock vapor case was used as the basis for the simulated observations. An example of these simulated data for the H$_2$O-rich case is shown in Figure \ref{fig:Sim_Obs} overplotted in comparison to the 1\% rock vapor case for both atmospheres respectively. The 1\% rock vapor cases were selected for the strong CO$_2$ absorption feature seen at 4.5\,$\mu$m in both spectra along with the silicate absorption feature seen in the O$_2$-rich atmosphere, serving as end-members of the species that could possibly be observed.

We then computed the $\chi^2$ values for the simulated observations relative to the blackbody spectrum. These values were used to determine the number of observations required to favor the 1\% rock-vapor atmosphere model over a featureless blackbody at the $3\sigma$ confidence level. Using only the NIRSpec portion of the spectrum, the 1\% rock vapor fraction composition was preferred at 3$\sigma$ for both the H$_2$O and O$_2$-rich atmospheres with only 1 observation. In contrast, it would require 5 and 
4 observations respectively to get the same confidence on the MIRI/LRS portion of the spectrum. This encourages the use of NIRSpec when planning future observations in order to best probe the potential CO$_2$ feature at 4.5 $\mu$m. 

TOI-431b has already been observed in a full-orbit phase curve using MIRI/LRS (program GO 8864, PI L. Dang). The phase curve will observe two additional eclipses, which, if they are consistent with our result, will increase the confidence of our tentative atmosphere detection and provide constraints on its day-night heat redistribution.

\section{Conclusions}
\label{sec:conc}
We observed a single secondary eclipse of the ultra-hot lava world TOI-431b with MIRI/LRS to probe its dayside temperature. We measure a secondary eclipse depth of $81 \pm 16$ ppm, corresponding to a brightness temperature ratio of 0.81 $\pm $0.10. We find that the eclipse depth is 1.8 $\sigma$ shallower than the theoretical maximum and potentially suggests either a high albedo and/or heat recirculation by an atmosphere. Given our current understanding of lava planetary surfaces, the dayside temperature likely cannot be fully explained by surface conditions alone and requires some contribution from an atmosphere. Our observation is only a single partial eclipse, but it is consistent with previous Spitzer observations of TOI-431b \citep{Monaghan2025}. A full orbit phase curve has been observed with MIRI/LRS (program GO 8864, PI L. Dang), and this observation could better characterize the nature of TOI-431b's potential atmosphere. Future observations at shorter wavelengths and further modeling efforts to understand lava planet atmospheres would enhance our understanding of TOI-431b.

\label{sec:conc}

\begin{acknowledgments} 
This work is based on observations with the NASA/ESA/CSA JWST. The data were obtained from the Mikulski Archive for Space Telescopes at the Space Telescope Science Institute (STScI), which is operated by the Association of Universities for Research in Astronomy, Incorporated, under NASA contract NAS5-03127. These observations are associated with program  JWST GO 4818. Support for this program was provided through a grant from STScI. All the {\it JWST} data used in this paper can be found in MAST: \dataset[10.17909/g5fa-6q07]{http://dx.doi.org/10.17909/g5fa-6q07}. We acknowledge the University of Maryland supercomputing resources (\url{http://hpcc.umd.edu}) made available for conducting the research reported in this paper.
R.L. is funded by the European Union (ERC, THIRSTEE, 101164189) and acknowledges financial support from the Severo Ochoa grant CEX2021-001131-S funded by MCIN/AEI/10.13039/501100011033.
D.K. was supported by NSFC grant number 12473064. A.A.A.P. acknowledges funding from a UK Science and Technology Facilities Council (STFC) Small Award, grant number UKRI/ST/B001171/1.
We acknowledge financial support from the Agencia Estatal de Investigaci\'on of the Ministerio de Ciencia e Innovaci\'on MCIN/AEI/10.13039/501100011033 and the ERDF “A way of making Europe” through projects PID2021-125627OB-C32 and PID2024-158486OB-C32. This work is supported by the European Union (ERC AdvG SPEAR, GA 101200674). Views and opinions expressed are however those of the authors only and do not necessarily reflect those of the European Union or the European Research Council. Neither the European Union nor the granting authority can be held responsible for them.

We thank the anonymous referee for their insightful comments and suggestions, which greatly improved the manuscript.
 \end{acknowledgments}

\appendix

\onecolumngrid

\section{\texttt{juliet} fit details}

In Table \ref{tab:posteriors}, we report the full priors and posteriors used in our new joint fits of available TESS and RV data. We also report select derived parameters using the stellar constraints of \citet{Osborn2021} in Table \ref{table:params_derived}. 

\begin{figure}[b!]
    \centering
    \includegraphics[width=0.8\linewidth]{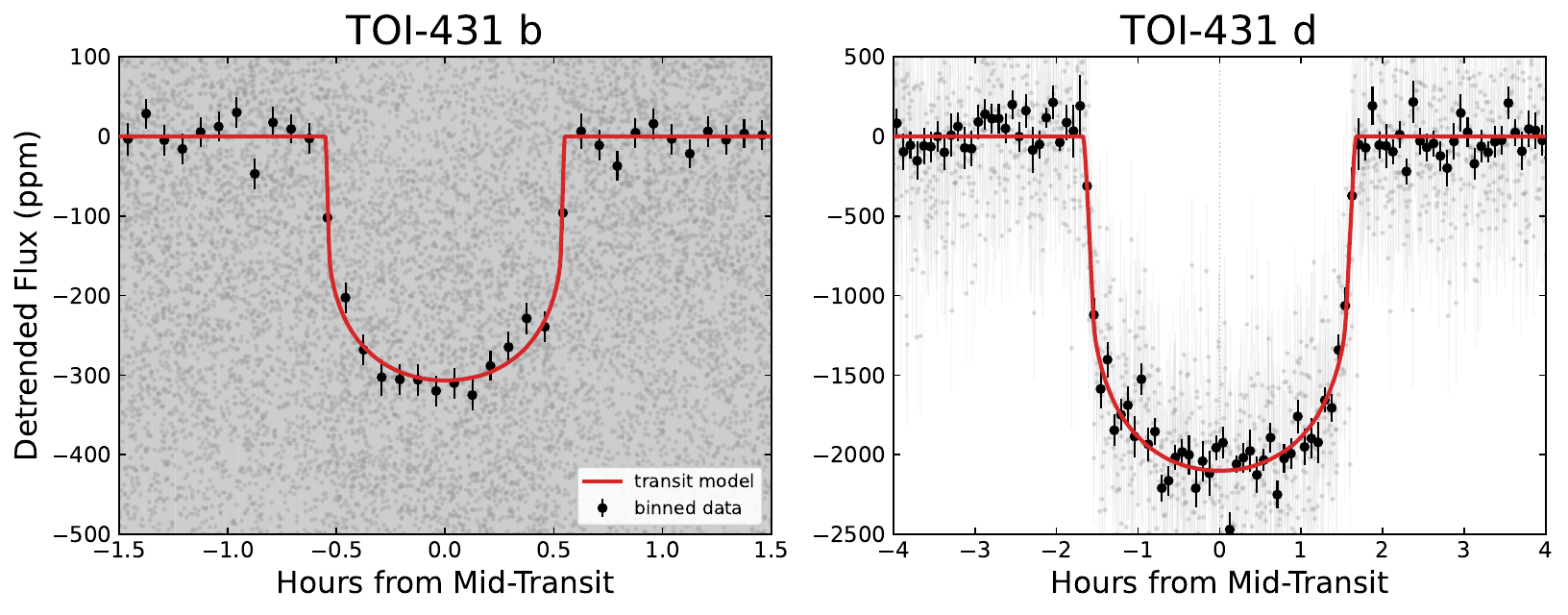}
    \caption{Detrended TESS data for the TOI-431 system, including the best-fit transit models (red lines) for planets b and d.  The raw data (grey) are binned at 5 minute resolution (black).}
    \label{fig:transits}
\end{figure}

\begin{figure}[b!]
    \centering
    \includegraphics[width=0.8\linewidth]{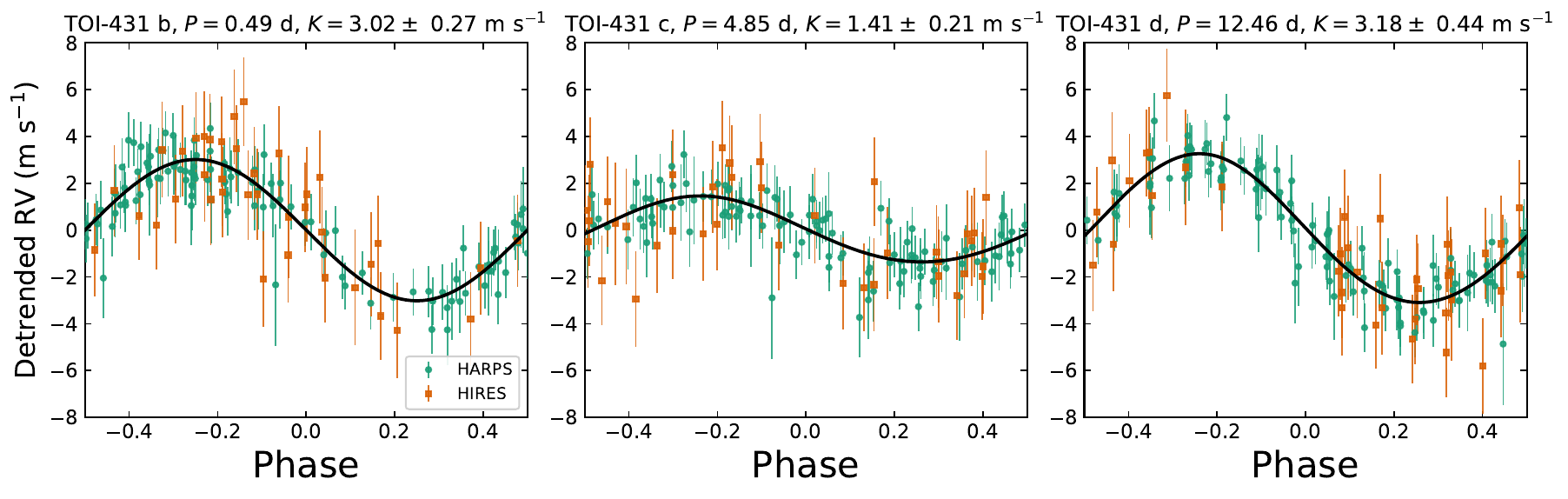}
    \caption{Detrended RV data for the TOI-431 system (green and orange points), including the best Keplerian fits for planets b, c, and d (black lines).}
    \label{fig:RVs}
\end{figure}

We show the detrended TESS transits of TOI-431~b and TOI-431~d in Figure \ref{fig:transits}, as well as the phase-folded RV data for each planet in Figure \ref{fig:RVs}. 

\begin{table*}[h!]
    \centering
    \caption{Priors and posterior distributions (median and 68\% credibility intervals) for each fit parameter of the final joint model obtained for the TOI-431 system using \texttt{juliet}. }
    \label{tab:posteriors}
\begin{tabular}{llcr}
\hline \hline
Parameter & Prior & Posterior & Units \\
\hline
$P_\mathrm{b}$      & $\mathcal{N}(0.4900,0.0005)$ & $0.490054945^{+0.000000079}_{-0.000000075}$ & d \\
$t_{0,\mathrm{b}}$  & $\mathcal{N}(2459199.43,0.04)$ & $2459199.4354\pm0.0002$ & BJD \\
$e_\mathrm{b}$      & 0 (fixed) & \dots & \dots \\
$\omega_\mathrm{b}$ & 90 (fixed) & \dots & $\deg$ \\
$b_{\mathrm{b}}$  & $\mathcal{U}(0, 1)$ & ${0.31}^{+{0.11}}_{-{0.12}}$ & \dots \\
$p_{\mathrm{b}}$  & $\mathcal{U}(0.01, 0.05)$ & $0.0161\pm0.0002$ & \dots \\
$K_\mathrm{b}$      & $\mathcal{U}(0, 20)$ & $3.02\pm0.27$ & $\mathrm{m\,s^{-1}}$ \\
\hline
$P_\mathrm{c}$      & $\mathcal{N}(4.85,0.01)$ & $4.853\pm0.005$ & d \\
$t_{0,\mathrm{c}}$  & $\mathcal{N}(2458625.85,0.5)$ & $2458625.74^{+0.16}_{-0.17}$ & BJD \\
$e_\mathrm{c}$      & $\mathcal{B}(1.52,29.0)$ & ${0.038}^{+{0.037}}_{-{0.022}}$ & \dots \\
$\omega_\mathrm{c}$ & $\mathcal{U}(-180,180)$ & ${-3}^{+113}_{-{117}}$ & $\deg$ \\
$K_\mathrm{c}$      & $\mathcal{U}(0, 20)$ & $1.41^{+0.21}_{-0.20}$ & $\mathrm{m\,s^{-1}}$ \\
\hline
$P_\mathrm{d}$      & $\mathcal{N}(12.4610,0.0005)$ & $12.461001\pm0.000003$ & d \\
$t_{0,\mathrm{d}}$  & $\mathcal{U}(2459188.29,0.04)$ & $2459188.2911\pm0.0003$ & BJD \\
$e_\mathrm{d}$      & $\mathcal{B}(1.52,29.0)$ & ${0.028}^{+{0.025}}_{-{0.016}}$ & \dots \\
$\omega_\mathrm{d}$ & $\mathcal{U}(-180,180)$ & ${9}^{+{93}}_{-{88}}$ & $\deg$ \\
$b_{\mathrm{d}}$  & $\mathcal{U}(0, 1)$ & ${0.16}^{+{0.15}}_{-{0.11}}$ & \dots \\
$p_{\mathrm{d}}$  & $\mathcal{U}(0.01, 0.05)$ & ${0.0418}^{+{0.0005}}_{-{0.0004}}$ & \dots \\
$K_\mathrm{d}$      & $\mathcal{U}(0, 20)$ & $3.18^{+0.40}_{-0.41}$ & $\mathrm{m\,s^{-1}}$ \\
\hline
$\rho_{\star}$ & $\mathcal{N}(2810,550)$ & $2964^{+267}_{-321}$ & kg\,m$^{-3}$ \\
$q_{1}$ & $\mathcal{N}(0.37,0.10)$ & $0.39^{+0.07}_{-0.06}$ & \dots \\
$q_{2}$ & $\mathcal{N}(0.34,0.10)$ & $0.34\pm0.07$ &  \dots\\
\hline
$\sigma_{\mathrm{TESS-S5+S6}}$     & $\mathcal{J}(10, 1000)$ & $172\pm4$ & ppm \\
$\sigma_{\mathrm{TESS-S32}}$    & $\mathcal{J}(10, 1000)$ & $161\pm6$ & ppm \\
$\sigma_{\mathrm{TESS-S98}}$    & $\mathcal{J}(10, 1000)$ & $302\pm3$ & ppm \\
$\sigma_{GP,\mathrm{TESS-S5+S6}}$     & $\mathcal{J}(1, 1000000)$ & $161^{+14}_{-13}$ & ppm \\
$\rho_{GP,\mathrm{TESS-S5+S6}}$     & $\mathcal{J}(0.01, 100)$ & $0.43^{+0.09}_{-0.06}$ & d \\
$\sigma_{GP,\mathrm{TESS-S32}}$    & $\mathcal{J}(1, 1000000)$ & $271^{+37}_{-29}$ & ppm \\
$\rho_{GP,\mathrm{TESS-S32}}$     & $\mathcal{J}(0.01, 100)$ & $0.53^{+0.11}_{-0.09}$ & d \\
$\sigma_{GP,\mathrm{TESS-S98}}$    & $\mathcal{J}(1, 1000000)$ & $2024^{+354}_{-278}$  & ppm \\
$\rho_{GP,\mathrm{TESS-S98}}$     & $\mathcal{J}(0.01, 100)$ & $1.94^{+0.26}_{-0.22}$ & d \\
\hline
$\gamma_{\mathrm{HARPS}}$ & $\mathcal{U}(-50.0,50.0)$ & $-2.4\pm3.2$ & $\mathrm{m\,s^{-1}}$ \\
$\sigma_{\mathrm{HARPS}}$ & $\mathcal{U}(0, 20)$ & ${0.19}^{+{0.19}}_{-{0.13}}$ & $\mathrm{m\,s^{-1}}$ \\
$\gamma_{\mathrm{HIRES}}$ & $\mathcal{U}(-50.0,50.0)$ & ${-1.6}^{+{3.6}}_{-{3.7}}$ & $\mathrm{m\,s^{-1}}$ \\
$\sigma_{\mathrm{HIRES}}$ & $\mathcal{U}(0, 20)$ & $1.6\pm0.4$ & $\mathrm{m\,s^{-1}}$ \\
$B_\mathrm{GP,RV}$   & $\mathcal{J}(10^{-2},10^{5})$    & ${44}^{+{33}}_{-{15}}$ & $\mathrm{m\,s^{-1}}$  \\
$C_\mathrm{GP,RV}$   & $\mathcal{J}(10^{-6},10^{-2})$   & ${0.00009}^{+{0.00195}}_{-{0.00008}}$ & \dots  \\
$L_\mathrm{GP,RV}$   & $\mathcal{J}(10^{-2},10^{5})$    & ${81}^{+{77}}_{-{35}}$ & \dots  \\
$P_\mathrm{rot,GP,RV}$   & $\mathcal{N}(30.5,0.7)$      & $30.1\pm0.7$ & d  \\
\hline
\end{tabular}
\tablecomments{
    The prior labels of $\mathcal{N}$, $\mathcal{U}$, $\mathcal{B}$, and $\mathcal{J}$ represent normal, uniform, beta, and Jeffrey's distributions, respectively \citep{Espinoza2019}
}
\end{table*}

\begin{table*}[h!]
    \centering
    \caption{Derived planetary parameters obtained for the TOI-431 system using the posterior values from the joint fit in Table~\ref{tab:posteriors} and stellar parameters from \citet{Osborn2021}. }
    \label{table:params_derived}
\begin{tabular}{ccccc}
\hline \hline
Parameter & Planet b & Planet c & Planet d & Units \\
\hline
$a/R_{\star}$ & $3.35^{+0.10}_{-0.13}$ & $15.5^{+0.5}_{-0.6}$ & $29.0^{+0.8}_{-1.1}$ & \dots \\
$i_\text{p}$ & $84.7^{+2.2}_{-2.1}$ & \dots & ${89.7}^{+{0.2}}_{-{0.3}}$ & deg \\
$M_\text{p}$ & ${3.16}^{+{0.34}}_{-{0.33}}$ & \dots & ${9.73}^{+{1.41}}_{-{1.35}}$ & $M_\oplus$ \\
$M_\text{p}\sin i$ & \dots & ${3.15}^{+{0.51}}_{-{0.49}}$ & \dots & $M_\oplus$ \\
$R_\text{p}$ & ${1.286}^{+{0.043}}_{-{0.042}}$ & \dots & $3.34\pm0.11$ & $R_\oplus$ \\
$\rho_\text{p}$ & ${8.2}^{+{1.3}}_{-{1.1}}$ & \dots & ${1.4}^{+{0.3}}_{-{0.2}}$ & \si{\gram\per\centi\meter\cubed} \\
$g_\text{p}$ & ${18.7}^{+{2.5}}_{-{2.3}}$ & \dots & ${8.5}^{+{1.4}}_{-{1.3}}$ & \si{\meter\per\second\squared} \\
$a_\text{p}$ & $0.0112\pm0.0003$ & ${0.0516}^{+{0.0015}}_{-{0.0016}}$ & ${0.097\pm0.003}$ & \si{\astronomicalunit} \\
$T_\textnormal{eq, p}$ & ${1875}^{+{45}}_{-{39}}$ & ${873}^{+21}_{-18}$ & ${638}^{+15}_{-13}$ & \si{\kelvin} \\
$T_{14}$ & $1.10\pm0.01$ & \dots & $3.36^{+0.09}_{-0.07}$ & hr \\
\hline
\end{tabular}
\tablecomments{Equilibrium temperature computed assuming zero Bond albedo and perfect energy redistribution.}
\end{table*}

\clearpage 

\bibliography{References_SMITH}

@article{horne1986optimal,
  title={An optimal extraction algorithm for CCD spectroscopy.},
  author={Horne, Keith},
  journal={Publications of the Astronomical Society of the Pacific},
  volume={98},
  number={604},
  pages={609},
  year={1986},
  publisher={IOP Publishing}
}

@ARTICLE{Kite2016,
       author = {{Kite}, Edwin S. and {Fegley}, Jr., Bruce and {Schaefer}, Laura and {Gaidos}, Eric},
        title = "{Atmosphere-interior Exchange on Hot, Rocky Exoplanets}",
      journal = {\apj},
         year = 2016,
        month = sep,
       volume = {828},
       number = {2},
          eid = {80},
        pages = {80},
          doi = {10.3847/0004-637X/828/2/80},
archivePrefix = {arXiv},
       eprint = {1606.06740},
 primaryClass = {astro-ph.EP},
       adsurl = {https://ui.adsabs.harvard.edu/abs/2016ApJ...828...80K}
}

@article{bell2024nightside,
  title={Nightside clouds and disequilibrium chemistry on the hot Jupiter WASP-43b},
  author={Bell, Taylor J and Crouzet, Nicolas and Cubillos, Patricio E and Kreidberg, Laura and Piette, Anjali AA and Roman, Michael T and Barstow, Joanna K and Blecic, Jasmina and Carone, Ludmila and Coulombe, Louis-Philippe and others},
  journal={Nature Astronomy},
  volume={8},
  number={7},
  pages={879--898},
  year={2024},
  publisher={Nature Publishing Group UK London}
}

@article{zhang2024gj,
  title={GJ 367b is a dark, hot, airless sub-Earth},
  author={Zhang, Michael and Hu, Renyu and Inglis, Julie and Dai, Fei and Bean, Jacob L and Knutson, Heather A and Lam, Kristine and Goffo, Elisa and Gandolfi, Davide},
  journal={The Astrophysical Journal Letters},
  volume={961},
  number={2},
  pages={L44},
  year={2024},
  publisher={IOP Publishing}
}

@article{speagle2020dynesty,
  title={dynesty: a dynamic nested sampling package for estimating Bayesian posteriors and evidences},
  author={Speagle, Joshua S},
  journal={Monthly Notices of the Royal Astronomical Society},
  volume={493},
  number={3},
  pages={3132--3158},
  year={2020},
  publisher={Oxford University Press}
}

@article{xue2025jwst,
  title={The JWST Rocky Worlds DDT Program reveals GJ 3929b to likely be a bare rock},
  author={Xue, Qiao and Zhang, Michael and Coy, Brandon Park and Brady, Madison and Ji, Xuan and Bean, Jacob L and Radica, Michael and Seifahrt, Andreas and St{\"u}rmer, Julian and Luque, Rafael and others},
  journal={The Astrophysical Journal Letters},
  volume={995},
  number={2},
  pages={L52},
  year={2025},
  publisher={IOP Publishing}
}

@article{seager2003unique,
  title={A unique solution of planet and star parameters from an extrasolar planet transit light curve},
  author={Seager, Sara and Mallen-Ornelas, Gabriela},
  journal={The Astrophysical Journal},
  volume={585},
  number={2},
  pages={1038--1055},
  year={2003}
}

@article{monaghan2026uniform,
  title={Uniform Reinterpretation of Rocky Exoplanet Secondary Eclipse Observations and the Impact of Stellar and Orbital Uncertainties},
  author={Monaghan, Christopher and Benneke, Bj{\"o}rn and Connors, Nicholas J and Coulombe, Louis-Philippe and Roy, Pierre-Alexis},
  journal={The Astronomical Journal},
  volume={171},
  number={5},
  pages={319},
  year={2026},
  publisher={The American Astronomical Society}
}

@article{coy2025population,
  title={Population-level Hypothesis Testing with Rocky Planet Emission Data: A Tentative Trend in the Brightness Temperatures of M-Earths},
  author={Coy, Brandon Park and Ih, Jegug and Kite, Edwin S and Koll, Daniel DB and Tenthoff, Moritz and Bean, Jacob L and Mansfield, Megan Weiner and Zhang, Michael and Xue, Qiao and Kempton, Eliza M-R and others},
  journal={The Astrophysical Journal},
  volume={987},
  number={1},
  pages={22},
  year={2025},
  publisher={IOP Publishing}
}

@ARTICLE{Foreman-Mackey2013,
       author = {{Foreman-Mackey}, Daniel and {Hogg}, David W. and {Lang}, Dustin and {Goodman}, Jonathan},
        title = "{emcee: The MCMC Hammer}",
      journal = {\pasp},
         year = 2013,
        month = mar,
       volume = {125},
       number = {925},
        pages = {306},
          doi = {10.1086/670067},
archivePrefix = {arXiv},
       eprint = {1202.3665},
 primaryClass = {astro-ph.IM},
       adsurl = {https://ui.adsabs.harvard.edu/abs/2013PASP..125..306F}
}

@ARTICLE{Bell:2022,
       author = {{Bell}, Taylor and {Ahrer}, Eva-Maria and {Brande}, Jonathan and {Carter}, Aarynn and {Feinstein}, Adina and {Caloca}, Giannina and {Mansfield}, Megan and {Zieba}, Sebastian and {Piaulet}, Caroline and {Benneke}, Bj{\"o}rn and {Filippazzo}, Joseph and {May}, Erin and {Roy}, Pierre-Alexis and {Kreidberg}, Laura and {Stevenson}, Kevin},
        title = "{Eureka!: An End-to-End Pipeline for JWST Time-Series Observations}",
      journal = {The Journal of Open Source Software},
         year = 2022,
        month = nov,
       volume = {7},
       number = {79},
          eid = {4503},
        pages = {4503},
          doi = {10.21105/joss.04503},
archivePrefix = {arXiv},
       eprint = {2207.03585},
 primaryClass = {astro-ph.IM},
       adsurl = {https://ui.adsabs.harvard.edu/abs/2022JOSS....7.4503B}
}

@article{Kreidberg_2015,
   title={<tt>batman</tt>: BAsic Transit Model cAlculatioN in Python},
   volume={127},
   ISSN={1538-3873},
   url={http://dx.doi.org/10.1086/683602},
   DOI={10.1086/683602},
   number={957},
   journal={Publications of the Astronomical Society of the Pacific},
   publisher={IOP Publishing},
   author={Kreidberg, Laura},
   year={2015},
   month=nov, pages={1161–1165} }

@ARTICLE{Cowan&Agol2011,
       author = {{Cowan}, Nicolas B. and {Agol}, Eric},
        title = "{The Statistics of Albedo and Heat Recirculation on Hot Exoplanets}",
      journal = {\apj},
         year = 2011,
        month = mar,
       volume = {729},
       number = {1},
          eid = {54},
        pages = {54},
          doi = {10.1088/0004-637X/729/1/54},
archivePrefix = {arXiv},
       eprint = {1001.0012},
 primaryClass = {astro-ph.EP},
       adsurl = {https://ui.adsabs.harvard.edu/abs/2011ApJ...729...54C}
}

@ARTICLE{Koll2019,
       author = {{Koll}, Daniel D.~B. and {Malik}, Matej and {Mansfield}, Megan and {Kempton}, Eliza M. -R. and {Kite}, Edwin and {Abbot}, Dorian and {Bean}, Jacob L.},
        title = "{Identifying Candidate Atmospheres on Rocky M Dwarf Planets via Eclipse Photometry}",
      journal = {\apj},
         year = 2019,
        month = dec,
       volume = {886},
       number = {2},
          eid = {140},
        pages = {140},
          doi = {10.3847/1538-4357/ab4c91},
archivePrefix = {arXiv},
       eprint = {1907.13138},
 primaryClass = {astro-ph.EP},
       adsurl = {https://ui.adsabs.harvard.edu/abs/2019ApJ...886..140K}
}

@ARTICLE{Mansfield2019,
       author = {{Mansfield}, Megan and {Kite}, Edwin S. and {Hu}, Renyu and {Koll}, Daniel D.~B. and {Malik}, Matej and {Bean}, Jacob L. and {Kempton}, Eliza M. -R.},
        title = "{Identifying Atmospheres on Rocky Exoplanets through Inferred High Albedo}",
      journal = {\apj},
         year = 2019,
        month = dec,
       volume = {886},
       number = {2},
          eid = {141},
        pages = {141},
          doi = {10.3847/1538-4357/ab4c90},
archivePrefix = {arXiv},
       eprint = {1907.13150},
 primaryClass = {astro-ph.EP},
       adsurl = {https://ui.adsabs.harvard.edu/abs/2019ApJ...886..141M}
}

@ARTICLE{Kempton2023,
       author = {{Kempton}, Eliza M. -R. and {Zhang}, Michael and {Bean}, Jacob L. and {Steinrueck}, Maria E. and {Piette}, Anjali A.~A. and {Parmentier}, Vivien and {Malsky}, Isaac and {Roman}, Michael T. and {Rauscher}, Emily and {Gao}, Peter and {Bell}, Taylor J. and {Xue}, Qiao and {Taylor}, Jake and {Savel}, Arjun B. and {Arnold}, Kenneth E. and {Nixon}, Matthew C. and {Stevenson}, Kevin B. and {Mansfield}, Megan and {Kendrew}, Sarah and {Zieba}, Sebastian and {Ducrot}, Elsa and {Dyrek}, Achr{\`e}ne and {Lagage}, Pierre-Olivier and {Stassun}, Keivan G. and {Henry}, Gregory W. and {Barman}, Travis and {Lupu}, Roxana and {Malik}, Matej and {Kataria}, Tiffany and {Ih}, Jegug and {Fu}, Guangwei and {Welbanks}, Luis and {McGill}, Peter},
        title = "{A reflective, metal-rich atmosphere for GJ 1214b from its JWST phase curve}",
      journal = {\nat},
         year = 2023,
        month = aug,
       volume = {620},
       number = {7972},
        pages = {67-71},
          doi = {10.1038/s41586-023-06159-5},
archivePrefix = {arXiv},
       eprint = {2305.06240},
 primaryClass = {astro-ph.EP},
       adsurl = {https://ui.adsabs.harvard.edu/abs/2023Natur.620...67K}
}

@ARTICLE{Greene2023,
       author = {{Greene}, Thomas P. and {Bell}, Taylor J. and {Ducrot}, Elsa and {Dyrek}, Achr{\`e}ne and {Lagage}, Pierre-Olivier and {Fortney}, Jonathan J.},
        title = "{Thermal emission from the Earth-sized exoplanet TRAPPIST-1 b using JWST}",
      journal = {\nat},
         year = 2023,
        month = jun,
       volume = {618},
       number = {7963},
        pages = {39-42},
          doi = {10.1038/s41586-023-05951-7},
archivePrefix = {arXiv},
       eprint = {2303.14849},
 primaryClass = {astro-ph.EP},
       adsurl = {https://ui.adsabs.harvard.edu/abs/2023Natur.618...39G}
}

@ARTICLE{Zieba2023,
       author = {{Zieba}, Sebastian and {Kreidberg}, Laura and {Ducrot}, Elsa and {Gillon}, Micha{\"e}l and {Morley}, Caroline and {Schaefer}, Laura and {Tamburo}, Patrick and {Koll}, Daniel D.~B. and {Lyu}, Xintong and {Acu{\~n}a}, Lorena and {Agol}, Eric and {Iyer}, Aishwarya R. and {Hu}, Renyu and {Lincowski}, Andrew P. and {Meadows}, Victoria S. and {Selsis}, Franck and {Bolmont}, Emeline and {Mandell}, Avi M. and {Suissa}, Gabrielle},
        title = "{No thick carbon dioxide atmosphere on the rocky exoplanet TRAPPIST-1 c}",
      journal = {\nat},
         year = 2023,
        month = aug,
       volume = {620},
       number = {7975},
        pages = {746-749},
          doi = {10.1038/s41586-023-06232-z},
archivePrefix = {arXiv},
       eprint = {2306.10150},
 primaryClass = {astro-ph.EP},
       adsurl = {https://ui.adsabs.harvard.edu/abs/2023Natur.620..746Z}
}

@INPROCEEDINGS{Allard2012,
       author = {{Allard}, France and {Homeier}, Derek and {Freytag}, Bernd},
        title = "{Stellar to Substellar Model Atmospheres}",
    booktitle = {From Interacting Binaries to Exoplanets: Essential Modeling Tools},
         year = 2012,
       editor = {{Richards}, Mercedes T. and {Hubeny}, Ivan},
       series = {IAU Symposium},
       volume = {282},
        month = apr,
        pages = {235-242},
          doi = {10.1017/S1743921311027438},
       adsurl = {https://ui.adsabs.harvard.edu/abs/2012IAUS..282..235A}
}

@misc{pysynphot2013,
  author       = {{STScI Development Team}},
  title        = {pysynphot: Synthetic Photometry Software Package},
  year         = {2013},
  publisher    = {Astrophysics Source Code Library},
  howpublished = {ascl:1303.023},
  note         = {Astrophysics Source Code Library, record ascl:1303.023}
}

@ARTICLE{Koll2016,
       author = {{Koll}, Daniel D.~B. and {Abbot}, Dorian S.},
        title = "{Temperature Structure and Atmospheric Circulation of Dry Tidally Locked Rocky Exoplanets}",
      journal = {\apj},
         year = 2016,
        month = jul,
       volume = {825},
       number = {2},
          eid = {99},
        pages = {99},
          doi = {10.3847/0004-637X/825/2/99},
archivePrefix = {arXiv},
       eprint = {1605.01066},
 primaryClass = {astro-ph.EP},
       adsurl = {https://ui.adsabs.harvard.edu/abs/2016ApJ...825...99K}
}

@ARTICLE{Dong2018,
       author = {{Dong}, Chuanfei and {Jin}, Meng and {Lingam}, Manasvi and {Airapetian}, Vladimir S. and {Ma}, Yingjuan and {van der Holst}, Bart},
        title = "{Atmospheric escape from the TRAPPIST-1 planets and implications for habitability}",
      journal = {Proceedings of the National Academy of Science},
         year = 2018,
        month = jan,
       volume = {115},
       number = {2},
        pages = {260-265},
          doi = {10.1073/pnas.1708010115},
archivePrefix = {arXiv},
       eprint = {1705.05535},
 primaryClass = {astro-ph.EP},
       adsurl = {https://ui.adsabs.harvard.edu/abs/2018PNAS..115..260D}
}

@ARTICLE{Kite&Barnnett2020,
       author = {{Kite}, Edwin S. and {Barnett}, Megan N.},
        title = "{Exoplanet secondary atmosphere loss and revival}",
      journal = {Proceedings of the National Academy of Science},
         year = 2020,
        month = jul,
       volume = {117},
        pages = {18264-18271},
          doi = {10.1073/pnas.2006177117},
archivePrefix = {arXiv},
       eprint = {2006.02589},
 primaryClass = {astro-ph.EP},
       adsurl = {https://ui.adsabs.harvard.edu/abs/2020PNAS..11718264K}
}

@ARTICLE{Schaefer2009,
       author = {{Schaefer}, Laura and {Fegley}, Bruce},
        title = "{Chemistry of Silicate Atmospheres of Evaporating Super-Earths}",
      journal = {\apjl},
         year = 2009,
        month = oct,
       volume = {703},
       number = {2},
        pages = {L113-L117},
          doi = {10.1088/0004-637X/703/2/L113},
archivePrefix = {arXiv},
       eprint = {0906.1204},
 primaryClass = {astro-ph.EP},
       adsurl = {https://ui.adsabs.harvard.edu/abs/2009ApJ...703L.113S}
}

@ARTICLE{Miguel2011,
       author = {{Miguel}, Y. and {Kaltenegger}, L. and {Fegley}, B. and {Schaefer}, L.},
        title = "{Compositions of Hot Super-earth Atmospheres: Exploring Kepler Candidates}",
      journal = {\apjl},
         year = 2011,
        month = dec,
       volume = {742},
       number = {2},
          eid = {L19},
        pages = {L19},
          doi = {10.1088/2041-8205/742/2/L19},
archivePrefix = {arXiv},
       eprint = {1110.2426},
 primaryClass = {astro-ph.EP},
       adsurl = {https://ui.adsabs.harvard.edu/abs/2011ApJ...742L..19M}
}

@ARTICLE{Espinoza2019,
       author = {{Espinoza}, N{\'e}stor and {Kossakowski}, Diana and {Brahm}, Rafael},
        title = "{juliet: a versatile modelling tool for transiting and non-transiting exoplanetary systems}",
      journal = {\mnras},
         year = "2019",
        month = "Dec",
       volume = {490},
       number = {2},
        pages = {2262-2283},
          doi = {10.1093/mnras/stz2688},
archivePrefix = {arXiv},
       eprint = {1812.08549},
 primaryClass = {astro-ph.EP},
       adsurl = {https://ui.adsabs.harvard.edu/abs/2019MNRAS.490.2262E}
}

@ARTICLE{Xue2024,
       author = {{Xue}, Qiao and {Bean}, Jacob L. and {Zhang}, Michael and {Mahajan}, Alexandra and {Ih}, Jegug and {Eastman}, Jason D. and {Lunine}, Jonathan and {Mansfield}, Megan Weiner and {Coy}, Brandon Park and {Kempton}, Eliza M.-R. and {Koll}, Daniel and {Kite}, Edwin},
        title = "{JWST Thermal Emission of the Terrestrial Exoplanet GJ 1132b}",
      journal = {\apjl},
         year = 2024,
        month = sep,
       volume = {973},
       number = {1},
          eid = {L8},
        pages = {L8},
          doi = {10.3847/2041-8213/ad72e9},
archivePrefix = {arXiv},
       eprint = {2408.13340},
 primaryClass = {astro-ph.EP},
       adsurl = {https://ui.adsabs.harvard.edu/abs/2024ApJ...973L...8X}
}

@ARTICLE{Monaghan2025,
       author = {{Monaghan}, Christopher and {Roy}, Pierre-Alexis and {Benneke}, Bj{\"o}rn and {Crossfield}, Ian J.~M. and {Coulombe}, Louis-Philippe and {Piaulet-Ghorayeb}, Caroline and {Kreidberg}, Laura and {Dressing}, Courtney D. and {Kane}, Stephen R. and {Dragomir}, Diana and {Werner}, Michael W. and {Parmentier}, Vivien and {Christiansen}, Jessie L. and {Morales}, Farisa Y. and {Berardo}, David and {Gorjian}, Varoujan},
        title = "{Low 4.5 {\ensuremath{\mu}}m Dayside Emission Disfavors a Dark Bare-rock Scenario for the Hot Super-Earth TOI-431 b}",
      journal = {\aj},
         year = 2025,
        month = may,
       volume = {169},
       number = {5},
          eid = {239},
        pages = {239},
          doi = {10.3847/1538-3881/adbe75},
archivePrefix = {arXiv},
       eprint = {2503.09698},
 primaryClass = {astro-ph.EP},
       adsurl = {https://ui.adsabs.harvard.edu/abs/2025AJ....169..239M}
}

@ARTICLE{Curry2025,
       author = {{Curry}, Alfred and {Mohanty}, Subhanjoy and {Owen}, James E.},
        title = "{Chemical evolution of an evaporating lava pool}",
      journal = {\mnras},
         year = 2025,
        month = jan,
       volume = {536},
       number = {1},
        pages = {913-930},
          doi = {10.1093/mnras/stae2583},
archivePrefix = {arXiv},
       eprint = {2411.13686},
 primaryClass = {astro-ph.EP},
       adsurl = {https://ui.adsabs.harvard.edu/abs/2025MNRAS.536..913C}
}

@ARTICLE{Osborn2021,
       author = {{Osborn}, Ares and {Armstrong}, David J. and {Cale}, Bryson and {Brahm}, Rafael and {Wittenmyer}, Robert A. and {Dai}, Fei and {Crossfield}, Ian J.~M. and {Bryant}, Edward M. and {Adibekyan}, Vardan and {Cloutier}, Ryan and {Collins}, Karen A. and {Delgado Mena}, E. and {Fridlund}, Malcolm and {Hellier}, Coel and {Howell}, Steve B. and {King}, George W. and {Lillo-Box}, Jorge and {Otegi}, Jon and {Sousa}, S. and {Stassun}, Keivan G. and {Matthews}, Elisabeth C. and {Ziegler}, Carl and {Ricker}, George and {Vanderspek}, Roland and {Latham}, David W. and {Seager}, S. and {Winn}, Joshua N. and {Jenkins}, Jon M. and {Acton}, Jack S. and {Addison}, Brett C. and {Anderson}, David R. and {Ballard}, Sarah and {Barrado}, David and {Barros}, Susana C.~C. and {Batalha}, Natalie and {Bayliss}, Daniel and {Barclay}, Thomas and {Benneke}, Bj{\"o}rn and {Berberian}, John and {Bouchy}, Francois and {Bowler}, Brendan P. and {Brice{\~n}o}, C{\'e}sar and {Burke}, Christopher J. and {Burleigh}, Matthew R. and {Casewell}, Sarah L. and {Ciardi}, David and {Collins}, Kevin I. and {Cooke}, Benjamin F. and {Demangeon}, Olivier D.~S. and {D{\'\i}az}, Rodrigo F. and {Dorn}, C. and {Dragomir}, Diana and {Dressing}, Courtney and {Dumusque}, Xavier and {Espinoza}, N{\'e}stor and {Figueira}, P. and {Fulton}, Benjamin and {Furlan}, E. and {Gaidos}, E. and {Geneser}, C. and {Gill}, Samuel and {Goad}, Michael R. and {Gonzales}, Erica J. and {Gorjian}, Varoujan and {G{\"u}nther}, Maximilian N. and {Helled}, Ravit and {Henderson}, Beth A. and {Henning}, Thomas and {Hogan}, Aleisha and {Hojjatpanah}, Saeed and {Horner}, Jonathan and {Howard}, Andrew W. and {Hoyer}, Sergio and {Huber}, Dan and {Isaacson}, Howard and {Jenkins}, James S. and {Jensen}, Eric L.~N. and {Jord{\'a}n}, Andr{\'e}s and {Kane}, Stephen R. and {Kidwell}, Richard C. and {Kielkopf}, John and {Law}, Nicholas and {Lendl}, Monika and {Lund}, M. and {Matson}, Rachel A. and {Mann}, Andrew W. and {McCormac}, James and {Mengel}, Matthew W. and {Morales}, Farisa Y. and {Nielsen}, Louise D. and {Okumura}, Jack and {Osborn}, Hugh P. and {Petigura}, Erik A. and {Plavchan}, Peter and {Pollacco}, Don and {Quintana}, Elisa V. and {Raynard}, Liam and {Robertson}, Paul and {Rose}, Mark E. and {Roy}, Arpita and {Reefe}, Michael and {Santerne}, Alexandre and {Santos}, Nuno C. and {Sarkis}, Paula and {Schlieder}, J. and {Schwarz}, Richard P. and {Scott}, Nicholas J. and {Shporer}, Avi and {Smith}, A.~M.~S. and {Stibbard}, C. and {Stockdale}, Chris and {Str{\o}m}, Paul A. and {Twicken}, Joseph D. and {Tan}, Thiam-Guan and {Tanner}, A. and {Teske}, J. and {Tilbrook}, Rosanna H. and {Tinney}, C.~G. and {Udry}, Stephane and {Villase{\~n}or}, Jesus Noel and {Vines}, Jose I. and {Wang}, Sharon X. and {Weiss}, Lauren M. and {West}, Richard G. and {Wheatley}, Peter J. and {Wright}, Duncan J. and {Zhang}, Hui and {Zohrabi}, F.},
        title = "{TOI-431/HIP 26013: a super-Earth and a sub-Neptune transiting a bright, early K dwarf, with a third RV planet}",
      journal = {\mnras},
         year = 2021,
        month = oct,
       volume = {507},
       number = {2},
        pages = {2782-2803},
          doi = {10.1093/mnras/stab2313},
archivePrefix = {arXiv},
       eprint = {2108.02310},
 primaryClass = {astro-ph.EP},
       adsurl = {https://ui.adsabs.harvard.edu/abs/2021MNRAS.507.2782O}
}

@article{Petrov2007,
author = {Petrov, Vadim and Vorobyev, Aleksey},
year = {2007},
month = {01},
pages = {pp. 321-329},
title = {Spectral emissivity and radiance temperature plateau of self-supporting Al2O3 melt at rapid solidification},
volume = {v. 35/36},
journal = {High Temperatures- High Pressures},
doi = {10.1068/htjr141}
}

@ARTICLE{Dvurechensky1979,
       author = {{Dvurechensky}, A.~V. and {Petrov}, V.~A. and {Yu Reznik}, V.},
        title = "{Spectral emissivity and absorption coefficient of silica glass at extremely high temperatures in the semitransparent region}",
      journal = {Infrared Physics},
         year = 1979,
        month = aug,
       volume = {19},
       number = {3},
        pages = {465-469},
          doi = {10.1016/0020-0891(79)90060-5},
       adsurl = {https://ui.adsabs.harvard.edu/abs/1979InfPh..19..465D}
}

@ARTICLE{Hu2024,
       author = {{Hu}, Renyu and {Bello-Arufe}, Aaron and {Zhang}, Michael and {Paragas}, Kimberly and {Zilinskas}, Mantas and {van Buchem}, Christiaan and {Bess}, Michael and {Patel}, Jayshil and {Ito}, Yuichi and {Damiano}, Mario and {Scheucher}, Markus and {Oza}, Apurva V. and {Knutson}, Heather A. and {Miguel}, Yamila and {Dragomir}, Diana and {Brandeker}, Alexis and {Demory}, Brice-Olivier},
        title = "{A secondary atmosphere on the rocky exoplanet 55 Cancri e}",
      journal = {\nat},
         year = 2024,
        month = jun,
       volume = {630},
       number = {8017},
        pages = {609-612},
          doi = {10.1038/s41586-024-07432-x},
archivePrefix = {arXiv},
       eprint = {2405.04744},
 primaryClass = {astro-ph.EP},
       adsurl = {https://ui.adsabs.harvard.edu/abs/2024Natur.630..609H}
}

@ARTICLE{Zilinskas2023,
       author = {{Zilinskas}, M. and {Miguel}, Y. and {van Buchem}, C.~P.~A. and {Snellen}, I.~A.~G.},
        title = "{Observability of silicates in volatile atmospheres of super-Earths and sub-Neptunes. Exploring the edge of the evaporation desert}",
      journal = {\aap},
         year = 2023,
        month = mar,
       volume = {671},
          eid = {A138},
        pages = {A138},
          doi = {10.1051/0004-6361/202245521},
       adsurl = {https://ui.adsabs.harvard.edu/abs/2023A&A...671A.138Z}
}

@ARTICLE{Piette2023,
       author = {{Piette}, Anjali A.~A. and {Gao}, Peter and {Brugman}, Kara and {Shahar}, Anat and {Lichtenberg}, Tim and {Miozzi}, Francesca and {Driscoll}, Peter},
        title = "{Rocky Planet or Water World? Observability of Low-density Lava World Atmospheres}",
      journal = {\apj},
         year = 2023,
        month = sep,
       volume = {954},
       number = {1},
          eid = {29},
        pages = {29},
          doi = {10.3847/1538-4357/acdef2},
archivePrefix = {arXiv},
       eprint = {2306.10100},
 primaryClass = {astro-ph.EP},
       adsurl = {https://ui.adsabs.harvard.edu/abs/2023ApJ...954...29P}
}

@ARTICLE{Rackham2018,
       author = {{Rackham}, Benjamin V. and {Apai}, D{\'a}niel and {Giampapa}, Mark S.},
        title = "{The Transit Light Source Effect: False Spectral Features and Incorrect Densities for M-dwarf Transiting Planets}",
      journal = {\apj},
         year = 2018,
        month = feb,
       volume = {853},
       number = {2},
          eid = {122},
        pages = {122},
          doi = {10.3847/1538-4357/aaa08c},
archivePrefix = {arXiv},
       eprint = {1711.05691},
 primaryClass = {astro-ph.EP},
       adsurl = {https://ui.adsabs.harvard.edu/abs/2018ApJ...853..122R}
}

@ARTICLE{Zilinskas2022,
       author = {{Zilinskas}, M. and {van Buchem}, C.~P.~A. and {Miguel}, Y. and {Louca}, A. and {Lupu}, R. and {Zieba}, S. and {van Westrenen}, W.},
        title = "{Observability of evaporating lava worlds}",
      journal = {\aap},
         year = 2022,
        month = may,
       volume = {661},
          eid = {A126},
        pages = {A126},
          doi = {10.1051/0004-6361/202142984},
archivePrefix = {arXiv},
       eprint = {2202.04759},
 primaryClass = {astro-ph.EP},
       adsurl = {https://ui.adsabs.harvard.edu/abs/2022A&A...661A.126Z}
}

@ARTICLE{Maurice2024,
       author = {{Maurice}, M. and {Dasgupta}, R. and {Hassanzadeh}, P.},
        title = "{Volatile atmospheres of lava worlds}",
      journal = {\aap},
         year = 2024,
        month = aug,
       volume = {688},
          eid = {A47},
        pages = {A47},
          doi = {10.1051/0004-6361/202347749},
archivePrefix = {arXiv},
       eprint = {2405.09284},
 primaryClass = {astro-ph.EP},
       adsurl = {https://ui.adsabs.harvard.edu/abs/2024A&A...688A..47M}
}

@ARTICLE{Cubillos2024,
       author = {{Cubillos}, Patricio E.},
        title = "{Gen TSO: A General JWST Simulator for Exoplanet Times-series Observations}",
      journal = {\pasp},
         year = 2024,
        month = dec,
       volume = {136},
       number = {12},
          eid = {124501},
        pages = {124501},
          doi = {10.1088/1538-3873/ad8fd4},
archivePrefix = {arXiv},
       eprint = {2410.04856},
 primaryClass = {astro-ph.EP},
       adsurl = {https://ui.adsabs.harvard.edu/abs/2024PASP..136l4501C}
}

@ARTICLE{Kipping2013,
       author = {{Kipping}, D.~M.},
        title = "{Parametrizing the exoplanet eccentricity distribution with the beta  distribution.}",
      journal = {\mnras},
         year = 2013,
        month = jul,
       volume = {434},
        pages = {L51-L55},
          doi = {10.1093/mnrasl/slt075},
archivePrefix = {arXiv},
       eprint = {1306.4982},
 primaryClass = {astro-ph.EP},
       adsurl = {https://ui.adsabs.harvard.edu/abs/2013MNRAS.434L..51K}
}

@article{magic2015stagger,
  title={The Stagger-grid: A grid of 3D stellar atmosphere models-IV. Limb darkening coefficients},
  author={Magic, Zazralt and Chiavassa, Andrea and Collet, Remo and Asplund, Martin},
  journal={Astronomy \& Astrophysics},
  volume={573},
  pages={A90},
  year={2015},
  publisher={EDP Sciences}
}

@article{Grant2024,
  title = {ExoTiC-LD: thirty seconds to stellar limb-darkening coefficients},
  author = {David Grant and Hannah R. Wakeford},
  journal = {Journal of Open Source Software},
  publisher = {The Open Journal},
  year = {2024},
  volume = {9},
  number = {100},
  pages = {6816},
  doi = {10.21105/joss.06816},
  url = {https://doi.org/10.21105/joss.06816}
}

@article{zieba2026dark,
  title={The dark and featureless surface of rocky exoplanet LHS 3844 b from JWST mid-infrared spectroscopy},
  author={Zieba, Sebastian and Kreidberg, Laura and Coy, Brandon P and Bello-Arufe, Aaron and Paragas, Kimberly and Lyu, Xintong and Hu, Renyu and Iyer, Aishwarya and Kite, Edwin S and Koll, Daniel DB and others},
  journal={Nature Astronomy},
  pages={1--16},
  year={2026},
  publisher={Nature Publishing Group UK London}
}

@ARTICLE{zeng2019,
       author = {{Zeng}, Li and {Jacobsen}, Stein B. and {Sasselov}, Dimitar D. and {Petaev}, Michail I. and {Vanderburg}, Andrew and {Lopez-Morales}, Mercedes and {Perez-Mercader}, Juan and {Mattsson}, Thomas R. and {Li}, Gongjie and {Heising}, Matthew Z. and {Bonomo}, Aldo S. and {Damasso}, Mario and {Berger}, Travis A. and {Cao}, Hao and {Levi}, Amit and {Wordsworth}, Robin D.},
        title = "{Growth model interpretation of planet size distribution}",
      journal = {Proceedings of the National Academy of Science},
         year = 2019,
        month = may,
       volume = {116},
       number = {20},
        pages = {9723-9728},
          doi = {10.1073/pnas.1812905116},
archivePrefix = {arXiv},
       eprint = {1906.04253},
 primaryClass = {astro-ph.EP},
       adsurl = {https://ui.adsabs.harvard.edu/abs/2019PNAS..116.9723Z}
}

@article{celerite2,
   author = {{Foreman-Mackey}, D.},
    title = "{Scalable Backpropagation for Gaussian Processes using Celerite}",
  journal = {Research Notes of the American Astronomical Society},
     year = 2018,
    month = feb,
   volume = 2,
   number = 1,
    pages = {31},
      doi = {10.3847/2515-5172/aaaf6c},
   adsurl = {http://adsabs.harvard.edu/abs/2018RNAAS...2a..31F}
}

@ARTICLE{vanEylen2019,
       author = {{Van Eylen}, Vincent and {Albrecht}, Simon and {Huang}, Xu and {MacDonald}, Mariah G. and {Dawson}, Rebekah I. and {Cai}, Maxwell X. and {Foreman-Mackey}, Daniel and {Lundkvist}, Mia S. and {Silva Aguirre}, Victor and {Snellen}, Ignas and {Winn}, Joshua N.},
        title = "{The Orbital Eccentricity of Small Planet Systems}",
      journal = {\aj},
         year = 2019,
        month = feb,
       volume = {157},
       number = {2},
          eid = {61},
        pages = {61},
          doi = {10.3847/1538-3881/aaf22f},
archivePrefix = {arXiv},
       eprint = {1807.00549},
 primaryClass = {astro-ph.EP},
       adsurl = {https://ui.adsabs.harvard.edu/abs/2019AJ....157...61V}
}

@ARTICLE{Essack2020,
       author = {{Essack}, Zahra and {Seager}, Sara and {Pajusalu}, Mihkel},
        title = "{Low-albedo Surfaces of Lava Worlds}",
      journal = {\apj},
         year = 2020,
        month = aug,
       volume = {898},
       number = {2},
          eid = {160},
        pages = {160},
          doi = {10.3847/1538-4357/ab9cba},
archivePrefix = {arXiv},
       eprint = {2008.02789},
 primaryClass = {astro-ph.EP},
       adsurl = {https://ui.adsabs.harvard.edu/abs/2020ApJ...898..160E}
}

@ARTICLE{Gandhi2017,
       author = {{Gandhi}, Siddharth and {Madhusudhan}, Nikku},
        title = "{GENESIS: new self-consistent models of exoplanetary spectra}",
      journal = {\mnras},
         year = 2017,
        month = dec,
       volume = {472},
       number = {2},
        pages = {2334-2355},
          doi = {10.1093/mnras/stx1601},
archivePrefix = {arXiv},
       eprint = {1706.02302},
 primaryClass = {astro-ph.EP},
       adsurl = {https://ui.adsabs.harvard.edu/abs/2017MNRAS.472.2334G}
}

@ARTICLE{Rothman2010,
   author = {{Rothman}, L.~S. and {Gordon}, I.~E. and {Barber}, R.~J. and 
	{Dothe}, H. and {Gamache}, R.~R. and {Goldman}, A. and {Perevalov}, V.~I. and 
	{Tashkun}, S.~A. and {Tennyson}, J.},
    title = "{HITEMP, the high-temperature molecular spectroscopic database}",
  journal = {\jqsrt},
     year = 2010,
    month = oct,
   volume = 111,
    pages = {2139-2150},
      doi = {10.1016/j.jqsrt.2010.05.001},
   adsurl = {http://adsabs.harvard.edu/abs/2010JQSRT.111.2139R}
}

@article{Yurchenko2021,
    author = {Yurchenko, Sergei N and Tennyson, Jonathan and Syme, Anna-Maree and Adam, Ahmad Y and Clark, Victoria H J and Cooper, Bridgette and Dobney, C Pria and Donnelly, Shaun T E and Gorman, Maire N and Lynas-Gray, Anthony E and Meltzer, Thomas and Owens, Alec and Qu, Qianwei and Semenov, Mikhail and Somogyi, Wilfrid and Upadhyay, Apoorva and Wright, Samuel and Zapata Trujillo, Juan C},
    title = "{ExoMol line lists – XLIV. Infrared and ultraviolet line list for silicon monoxide (28Si16O)}",
    journal = {Monthly Notices of the Royal Astronomical Society},
    volume = {510},
    number = {1},
    pages = {903-919},
    year = {2021},
    month = {11},
    issn = {0035-8711},
    doi = {10.1093/mnras/stab3267},
    url = {https://doi.org/10.1093/mnras/stab3267},
    eprint = {https://academic.oup.com/mnras/article-pdf/510/1/903/41899259/stab3267.pdf},
}

@ARTICLE{Owens2020,
       author = {{Owens}, A. and {Conway}, E.~K. and {Tennyson}, J. and {Yurchenko}, S.~N.},
        title = "{ExoMol line lists - XXXVIII. High-temperature molecular line list of silicon dioxide (SiO$_{2}$)}",
      journal = {\mnras},
         year = 2020,
        month = jun,
       volume = {495},
       number = {2},
        pages = {1927-1933},
          doi = {10.1093/mnras/staa1287},
archivePrefix = {arXiv},
       eprint = {2005.13586},
 primaryClass = {astro-ph.EP},
       adsurl = {https://ui.adsabs.harvard.edu/abs/2020MNRAS.495.1927O}
}

@ARTICLE{Patrascu2015,
       author = {{Patrascu}, Andrei T. and {Yurchenko}, Sergei N. and {Tennyson}, Jonathan},
        title = "{ExoMol molecular line lists - IX. The spectrum of AlO}",
      journal = {\mnras},
         year = 2015,
        month = jun,
       volume = {449},
       number = {4},
        pages = {3613-3619},
          doi = {10.1093/mnras/stv507},
archivePrefix = {arXiv},
       eprint = {1504.02938},
 primaryClass = {astro-ph.GA},
       adsurl = {https://ui.adsabs.harvard.edu/abs/2015MNRAS.449.3613P}
}

@ARTICLE{Li2019,
       author = {{Li}, Heng Ying and {Tennyson}, Jonathan and {Yurchenko}, Sergei N.},
        title = "{ExoMol line lists - XXXII. The rovibronic spectrum of MgO}",
      journal = {\mnras},
         year = 2019,
        month = jun,
       volume = {486},
       number = {2},
        pages = {2351-2365},
          doi = {10.1093/mnras/stz912},
archivePrefix = {arXiv},
       eprint = {1904.12155},
 primaryClass = {astro-ph.SR},
       adsurl = {https://ui.adsabs.harvard.edu/abs/2019MNRAS.486.2351L}
}

@ARTICLE{Mitev2022,
       author = {{Mitev}, G.~B. and {Taylor}, S. and {Tennyson}, Jonathan and {Yurchenko}, S.~N. and {Buchachenko}, A.~A. and {Stolyarov}, A.~V.},
        title = "{ExoMol molecular line lists - XLIII. Rovibronic transitions corresponding to the close-lying X $^{2}${\ensuremath{\Pi}} and A $^{2}${\ensuremath{\Sigma}}$^{+}$ states of NaO}",
      journal = {\mnras},
         year = 2022,
        month = apr,
       volume = {511},
       number = {2},
        pages = {2349-2355},
          doi = {10.1093/mnras/stab3357},
archivePrefix = {arXiv},
       eprint = {2111.08424},
 primaryClass = {physics.chem-ph},
       adsurl = {https://ui.adsabs.harvard.edu/abs/2022MNRAS.511.2349M}
}

@ARTICLE{McKemmish2019,
       author = {{McKemmish}, Laura K. and {Masseron}, Thomas and {Hoeijmakers}, H. Jens and {P{\'e}rez-Mesa}, V{\'\i}ctor and {Grimm}, Simon L. and {Yurchenko}, Sergei N. and {Tennyson}, Jonathan},
        title = "{ExoMol molecular line lists - XXXIII. The spectrum of Titanium Oxide}",
      journal = {\mnras},
         year = 2019,
        month = sep,
       volume = {488},
       number = {2},
        pages = {2836-2854},
          doi = {10.1093/mnras/stz1818},
archivePrefix = {arXiv},
       eprint = {1905.04587},
 primaryClass = {astro-ph.SR},
       adsurl = {https://ui.adsabs.harvard.edu/abs/2019MNRAS.488.2836M}
}

@ARTICLE{Gordon2017,
       author = {{Gordon}, I.~E. and {Rothman}, L.~S. and {Hill}, C. and {Kochanov}, R.~V. and {Tan}, Y. and {Bernath}, P.~F. and {Birk}, M. and {Boudon}, V. and {Campargue}, A. and {Chance}, K.~V. and {Drouin}, B.~J. and {Flaud}, J. -M. and {Gamache}, R.~R. and {Hodges}, J.~T. and {Jacquemart}, D. and {Perevalov}, V.~I. and {Perrin}, A. and {Shine}, K.~P. and {Smith}, M. -A.~H. and {Tennyson}, J. and {Toon}, G.~C. and {Tran}, H. and {Tyuterev}, V.~G. and {Barbe}, A. and {Cs{\'a}sz{\'a}r}, A.~G. and {Devi}, V.~M. and {Furtenbacher}, T. and {Harrison}, J.~J. and {Hartmann}, J. -M. and {Jolly}, A. and {Johnson}, T.~J. and {Karman}, T. and {Kleiner}, I. and {Kyuberis}, A.~A. and {Loos}, J. and {Lyulin}, O.~M. and {Massie}, S.~T. and {Mikhailenko}, S.~N. and {Moazzen-Ahmadi}, N. and {M{\"u}ller}, H.~S.~P. and {Naumenko}, O.~V. and {Nikitin}, A.~V. and {Polyansky}, O.~L. and {Rey}, M. and {Rotger}, M. and {Sharpe}, S.~W. and {Sung}, K. and {Starikova}, E. and {Tashkun}, S.~A. and {Auwera}, J. Vander and {Wagner}, G. and {Wilzewski}, J. and {Wcis{\l}o}, P. and {Yu}, S. and {Zak}, E.~J.},
        title = "{The HITRAN2016 molecular spectroscopic database}",
      journal = {\jqsrt},
         year = 2017,
        month = dec,
       volume = {203},
        pages = {3-69},
          doi = {10.1016/j.jqsrt.2017.06.038},
       adsurl = {https://ui.adsabs.harvard.edu/abs/2017JQSRT.203....3G}
}

@ARTICLE{Dulick2003,
       author = {{Dulick}, M. and {Bauschlicher}, Jr., C.~W. and {Burrows}, Adam and {Sharp}, C.~M. and {Ram}, R.~S. and {Bernath}, Peter},
        title = "{Line Intensities and Molecular Opacities of the FeH F $^{4}${\ensuremath{\Delta}}$_{i}$-X $^{4}${\ensuremath{\Delta}}$_{i}$ Transition}",
      journal = {\apj},
         year = 2003,
        month = sep,
       volume = {594},
       number = {1},
        pages = {651-663},
          doi = {10.1086/376791},
archivePrefix = {arXiv},
       eprint = {astro-ph/0305162},
 primaryClass = {astro-ph},
       adsurl = {https://ui.adsabs.harvard.edu/abs/2003ApJ...594..651D}
}

@ARTICLE{Bernath2020,
       author = {{Bernath}, Peter F.},
        title = "{MoLLIST: Molecular Line Lists, Intensities and Spectra}",
      journal = {\jqsrt},
         year = 2020,
        month = jan,
       volume = {240},
          eid = {106687},
        pages = {106687},
          doi = {10.1016/j.jqsrt.2019.106687},
       adsurl = {https://ui.adsabs.harvard.edu/abs/2020JQSRT.24006687B}
}

@ARTICLE{Rivlin2015,
       author = {{Rivlin}, Tom and {Lodi}, Lorenzo and {Yurchenko}, Sergei N. and {Tennyson}, Jonathan and {Le Roy}, Robert J.},
        title = "{ExoMol molecular line lists - X. The spectrum of sodium hydride}",
      journal = {\mnras},
         year = 2015,
        month = jul,
       volume = {451},
       number = {1},
        pages = {634-638},
          doi = {10.1093/mnras/stv979},
archivePrefix = {arXiv},
       eprint = {1506.00174},
 primaryClass = {astro-ph.GA},
       adsurl = {https://ui.adsabs.harvard.edu/abs/2015MNRAS.451..634R}
}

@ARTICLE{Owens2021,
       author = {{Owens}, A. and {Tennyson}, J. and {Yurchenko}, S.~N.},
        title = "{ExoMol line lists - XLI. High-temperature molecular line lists for the alkali metal hydroxides KOH and NaOH}",
      journal = {\mnras},
         year = 2021,
        month = mar,
       volume = {502},
       number = {1},
        pages = {1128-1135},
          doi = {10.1093/mnras/staa4041},
archivePrefix = {arXiv},
       eprint = {2103.01601},
 primaryClass = {astro-ph.EP},
       adsurl = {https://ui.adsabs.harvard.edu/abs/2021MNRAS.502.1128O}
}

@ARTICLE{Grimm2021,
       author = {{Grimm}, Simon L. and {Malik}, Matej and {Kitzmann}, Daniel and {Guzm{\'a}n-Mesa}, Andrea and {Hoeijmakers}, H. Jens and {Fisher}, Chloe and {Mendon{\c{c}}a}, Jo{\~a}o M. and {Yurchenko}, Sergey N. and {Tennyson}, Jonathan and {Alesina}, Fabien and {Buchschacher}, Nicolas and {Burnier}, Julien and {Segransan}, Damien and {Kurucz}, Robert L. and {Heng}, Kevin},
        title = "{HELIOS-K 2.0 Opacity Calculator and Open-source Opacity Database for Exoplanetary Atmospheres}",
      journal = {\apjs},
         year = 2021,
        month = mar,
       volume = {253},
       number = {1},
          eid = {30},
        pages = {30},
          doi = {10.3847/1538-4365/abd773},
archivePrefix = {arXiv},
       eprint = {2101.02005},
 primaryClass = {astro-ph.EP},
       adsurl = {https://ui.adsabs.harvard.edu/abs/2021ApJS..253...30G}
}

@INPROCEEDINGS{Kurucz2018,
       author = {{Kurucz}, R.~L.},
        title = "{Including All the Lines: Data Releases for Spectra and Opacities through 2017}",
    booktitle = {Workshop on Astrophysical Opacities},
         year = 2018,
       series = {Astronomical Society of the Pacific Conference Series},
       volume = {515},
        month = aug,
        pages = {47},
       adsurl = {https://ui.adsabs.harvard.edu/abs/2018ASPC..515...47K}
}

@ARTICLE{Piette2020_GENESIS,
       author = {{Piette}, Anjali A.~A. and {Madhusudhan}, Nikku},
        title = "{On the Temperature Profiles and Emission Spectra of Mini-Neptune Atmospheres}",
      journal = {\apj},
         year = 2020,
        month = dec,
       volume = {904},
       number = {2},
          eid = {154},
        pages = {154},
          doi = {10.3847/1538-4357/abbfb1},
archivePrefix = {arXiv},
       eprint = {2009.11290},
 primaryClass = {astro-ph.EP},
       adsurl = {https://ui.adsabs.harvard.edu/abs/2020ApJ...904..154P}
}

@ARTICLE{Stock2022_fastchem,
       author = {{Stock}, Joachim W. and {Kitzmann}, Daniel and {Patzer}, A. Beate C.},
        title = "{FASTCHEM 2 : an improved computer program to determine the gas-phase chemical equilibrium composition for arbitrary element distributions}",
      journal = {\mnras},
         year = 2022,
        month = dec,
       volume = {517},
       number = {3},
        pages = {4070-4080},
          doi = {10.1093/mnras/stac2623},
archivePrefix = {arXiv},
       eprint = {2206.08247},
 primaryClass = {astro-ph.EP},
       adsurl = {https://ui.adsabs.harvard.edu/abs/2022MNRAS.517.4070S}
}

@article{Wolf2022_vaporock,
  title={VapoRock: Thermodynamics of vaporized silicate melts for modeling volcanic outgassing and magma ocean atmospheres},
  author={Wolf, Aaron S and J{\"a}ggi, Noah and Sossi, Paolo A and Bower, Dan J},
  journal={The Astrophysical Journal},
  volume={947},
  number={2},
  pages={64},
  year={2023},
  publisher={The American Astronomical Society}
}

@ARTICLE{Ducrot2025,
       author = {{Ducrot}, Elsa and {Lagage}, Pierre-Olivier and {Min}, Michiel and {Gillon}, Micha{\"e}l and {Bell}, Taylor J. and {Tremblin}, Pascal and {Greene}, Thomas and {Dyrek}, Achr{\`e}ne and {Bouwman}, Jeroen and {Waters}, Rens and {G{\"u}del}, Manuel and {Henning}, Thomas and {Vandenbussche}, Bart and {Absil}, Olivier and {Barrado}, David and {Boccaletti}, Anthony and {Coulais}, Alain and {Decin}, Leen and {Edwards}, Billy and {Gastaud}, Ren{\'e} and {Glasse}, Alistair and {Kendrew}, Sarah and {Olofsson}, Goran and {Patapis}, Polychronis and {Pye}, John and {Rouan}, Daniel and {Whiteford}, Niall and {Argyriou}, Ioannis and {Cossou}, Christophe and {Glauser}, Adrian M. and {Krause}, Oliver and {Lahuis}, Fred and {Royer}, Pierre and {Scheithauer}, Silvia and {Colina}, Luis and {van Dishoeck}, Ewine F. and {Ostlin}, G{\"o}ran and {Ray}, Tom P. and {Wright}, Gillian},
        title = "{Combined analysis of the 12.8 and 15 {\ensuremath{\mu}}m JWST/MIRI eclipse observations of TRAPPIST-1 b}",
      journal = {Nature Astronomy},
         year = 2025,
        month = mar,
       volume = {9},
        pages = {358-369},
          doi = {10.1038/s41550-024-02428-z},
archivePrefix = {arXiv},
       eprint = {2412.11627},
 primaryClass = {astro-ph.EP},
       adsurl = {https://ui.adsabs.harvard.edu/abs/2025NatAs...9..358D}
}

@INPROCEEDINGS{Wachiraphan2025,
       author = {{Wachiraphan}, Patcharapol and {Berta-Thompson}, Zachory and {Diamond-Lowe}, Hannah and {Winters}, Jennifer and {Murray}, Catriona and {Zhang}, Michael and {Xue}, Qiao and {Morley}, Caroline and {Rosario-Franco}, Marialis and {Duvvuri}, Girish},
        title = "{The Thermal Emission Spectrum of the Nearby Rocky Exoplanet LTT 1445A b from JWST MIRI/LRS}",
    booktitle = {American Astronomical Society Meeting Abstracts \#245},
         year = 2025,
       series = {American Astronomical Society Meeting Abstracts},
       volume = {245},
        month = jan,
          eid = {119.06},
        pages = {119.06},
       adsurl = {https://ui.adsabs.harvard.edu/abs/2025AAS...24511906W}
}

@ARTICLE{Fortune2025,
       author = {{Fortune}, Mark and {Gibson}, Neale P. and {Diamond-Lowe}, Hannah and {Mendon{\c{c}}a}, Jo{\~a}o M. and {Gressier}, Am{\'e}lie and {Kitzmann}, Daniel and {Allen}, Natalie H. and {August}, Prune C. and {Ih}, Jegug and {Meier Vald{\'e}s}, Erik and {Zgraggen}, Merlin and {Buchhave}, Lars A. and {Demory}, Brice-Olivier and {Espinoza}, N{\'e}stor and {Heng}, Kevin and {Jones}, Kathryn and {Rathcke}, Alexander D.},
        title = "{Hot Rocks Survey: III. A deep eclipse for LHS 1140c and a new Gaussian process method to account for correlated noise in individual pixels}",
      journal = {\aap},
         year = 2025,
        month = sep,
       volume = {701},
          eid = {A25},
        pages = {A25},
          doi = {10.1051/0004-6361/202554198},
archivePrefix = {arXiv},
       eprint = {2505.22186},
 primaryClass = {astro-ph.EP},
       adsurl = {https://ui.adsabs.harvard.edu/abs/2025A&A...701A..25F}
}

@ARTICLE{Lin2026,
       author = {{Lin}, Zifan and {Daylan}, Tansu},
        title = "{The Persistent Thermal Anomalies in Rocky Worlds}",
      journal = {arXiv e-prints},
         year = 2026,
        month = jan,
          eid = {arXiv:2601.00412},
        pages = {arXiv:2601.00412},
          doi = {10.48550/arXiv.2601.00412},
archivePrefix = {arXiv},
       eprint = {2601.00412},
 primaryClass = {astro-ph.EP},
       adsurl = {https://ui.adsabs.harvard.edu/abs/2026arXiv260100412L}
}

@ARTICLE{WeinerMansfield2024,
       author = {{Weiner Mansfield}, Megan and {Xue}, Qiao and {Zhang}, Michael and {Mahajan}, Alexandra S. and {Ih}, Jegug and {Koll}, Daniel and {Bean}, Jacob L. and {Coy}, Brandon Park and {Eastman}, Jason D. and {Kempton}, Eliza M.-R. and {Kite}, Edwin S.},
        title = "{No Thick Atmosphere on the Terrestrial Exoplanet Gl 486b}",
      journal = {\apjl},
         year = 2024,
        month = nov,
       volume = {975},
       number = {1},
          eid = {L22},
        pages = {L22},
          doi = {10.3847/2041-8213/ad8161},
archivePrefix = {arXiv},
       eprint = {2408.15123},
 primaryClass = {astro-ph.EP},
       adsurl = {https://ui.adsabs.harvard.edu/abs/2024ApJ...975L..22W}
}

@ARTICLE{Kriedberg2019,
       author = {{Kreidberg}, Laura and {Koll}, Daniel D.~B. and {Morley}, Caroline and {Hu}, Renyu and {Schaefer}, Laura and {Deming}, Drake and {Stevenson}, Kevin B. and {Dittmann}, Jason and {Vanderburg}, Andrew and {Berardo}, David and {Guo}, Xueying and {Stassun}, Keivan and {Crossfield}, Ian and {Charbonneau}, David and {Latham}, David W. and {Loeb}, Abraham and {Ricker}, George and {Seager}, Sara and {Vanderspek}, Roland},
        title = "{Absence of a thick atmosphere on the terrestrial exoplanet LHS 3844b}",
      journal = {\nat},
         year = 2019,
        month = aug,
       volume = {573},
       number = {7772},
        pages = {87-90},
          doi = {10.1038/s41586-019-1497-4},
archivePrefix = {arXiv},
       eprint = {1908.06834},
 primaryClass = {astro-ph.EP},
       adsurl = {https://ui.adsabs.harvard.edu/abs/2019Natur.573...87K}
}

@ARTICLE{Allen2025,
       author = {{Allen}, Natalie H. and {Espinoza}, N{\'e}stor and {Diamond-Lowe}, Hannah and {Mendon{\c{c}}a}, Jo{\~a}o M. and {Demory}, Brice-Olivier and {Gressier}, Am{\'e}lie and {Ih}, Jegug and {Fortune}, Mark and {August}, Prune C. and {Holmberg}, M{\r{a}}ns and {Meier Vald{\'e}s}, Erik and {Zgraggen}, Merlin and {Buchhave}, Lars A. and {Burgasser}, Adam J. and {Fisher}, Chloe and {Gibson}, Neale P. and {Heng}, Kevin and {Hoeijmakers}, Jens and {Kitzmann}, Daniel and {Prinoth}, Bibiana and {Rathcke}, Alexander D. and {Morris}, Brett M.},
        title = "{Hot Rocks Survey. IV. Emission from LTT 3780 b Is Consistent with a Bare Rock}",
      journal = {\aj},
         year = 2025,
        month = oct,
       volume = {170},
       number = {4},
          eid = {240},
        pages = {240},
          doi = {10.3847/1538-3881/adfc51},
archivePrefix = {arXiv},
       eprint = {2508.14210},
 primaryClass = {astro-ph.EP},
       adsurl = {https://ui.adsabs.harvard.edu/abs/2025AJ....170..240A}
}

@ARTICLE{Crossfield2022,
       author = {{Crossfield}, Ian J.~M. and {Malik}, Matej and {Hill}, Michelle L. and {Kane}, Stephen R. and {Foley}, Bradford and {Polanski}, Alex S. and {Coria}, David and {Brande}, Jonathan and {Zhang}, Yanzhe and {Wienke}, Katherine and {Kreidberg}, Laura and {Cowan}, Nicolas B. and {Dragomir}, Diana and {Gorjian}, Varoujan and {Mikal-Evans}, Thomas and {Benneke}, Bj{\"o}rn and {Christiansen}, Jessie L. and {Deming}, Drake and {Morales}, Farisa Y.},
        title = "{GJ 1252b: A Hot Terrestrial Super-Earth with No Atmosphere}",
      journal = {\apjl},
         year = 2022,
        month = sep,
       volume = {937},
       number = {1},
          eid = {L17},
        pages = {L17},
          doi = {10.3847/2041-8213/ac886b},
archivePrefix = {arXiv},
       eprint = {2208.09479},
 primaryClass = {astro-ph.EP},
       adsurl = {https://ui.adsabs.harvard.edu/abs/2022ApJ...937L..17C}
}

@ARTICLE{Luque2025,
       author = {{Luque}, Rafael and {Coy}, Brandon Park and {Xue}, Qiao and {Feinstein}, Adina D. and {Ahrer}, Eva-Maria and {Changeat}, Quentin and {Zhang}, Michael and {Moran}, Sarah E. and {Bean}, Jacob L. and {Kite}, Edwin and {Weiner Mansfield}, Megan and {Pall{\'e}}, Enric},
        title = "{A Dark, Bare Rock for TOI-1685 b from a JWST NIRSpec G395H Phase Curve}",
      journal = {\aj},
         year = 2025,
        month = jul,
       volume = {170},
       number = {1},
          eid = {49},
        pages = {49},
          doi = {10.3847/1538-3881/addb40},
archivePrefix = {arXiv},
       eprint = {2412.03411},
 primaryClass = {astro-ph.EP},
       adsurl = {https://ui.adsabs.harvard.edu/abs/2025AJ....170...49L}
}

@ARTICLE{Zieba2022,
       author = {{Zieba}, S. and {Zilinskas}, M. and {Kreidberg}, L. and {Nguyen}, T.~G. and {Miguel}, Y. and {Cowan}, N.~B. and {Pierrehumbert}, R. and {Carone}, L. and {Dang}, L. and {Hammond}, M. and {Louden}, T. and {Lupu}, R. and {Malavolta}, L. and {Stevenson}, K.~B.},
        title = "{K2 and Spitzer phase curves of the rocky ultra-short-period planet K2-141 b hint at a tenuous rock vapor atmosphere}",
      journal = {\aap},
         year = 2022,
        month = aug,
       volume = {664},
          eid = {A79},
        pages = {A79},
          doi = {10.1051/0004-6361/202142912},
archivePrefix = {arXiv},
       eprint = {2203.00370},
 primaryClass = {astro-ph.EP},
       adsurl = {https://ui.adsabs.harvard.edu/abs/2022A&A...664A..79Z}
}

@ARTICLE{Teske2025,
       author = {{Teske}, Johanna K. and {Wallack}, Nicole L. and {Piette}, Anjali A.~A. and {Dang}, Lisa and {Lichtenberg}, Tim and {Plotnykov}, Mykhaylo and {Pierrehumbert}, Raymond and {Postolec}, Emma and {Boucher}, Samuel and {McGinty}, Alex and {Peng}, Bo and {Valencia}, Diana and {Hammond}, Mark},
        title = "{A Thick Volatile Atmosphere on the Ultrahot Super-Earth TOI-561 b}",
      journal = {\apjl},
         year = 2025,
        month = dec,
       volume = {995},
       number = {2},
          eid = {L39},
        pages = {L39},
          doi = {10.3847/2041-8213/ae0a4c},
archivePrefix = {arXiv},
       eprint = {2509.17231},
 primaryClass = {astro-ph.EP},
       adsurl = {https://ui.adsabs.harvard.edu/abs/2025ApJ...995L..39T}
}

@article{Coy2026,
  title={Evidence for an Atmosphere on the Ultra-short-period Super-Earth HD 3167 b},
  author={Coy, Brandon Park and Xue, Qiao and Weiner Mansfield, Megan and Eastman, Jason D and Piette, Anjali AA and Fairnington, Tyler and Smith, Cole and Zhang, Michael and Kempton, Eliza M-R and Bean, Jacob L and others},
  journal={The Astrophysical Journal Letters},
  volume={1005},
  number={2},
  pages={L77},
  year={2026},
  publisher={The American Astronomical Society}
}

@ARTICLE{Xue2025,
       author = {{Xue}, Qiao and {Zhang}, Michael and {Coy}, Brandon Park and {Brady}, Madison and {Ji}, Xuan and {Bean}, Jacob L. and {Radica}, Michael and {Seifahrt}, Andreas and {St{\"u}rmer}, Julian and {Luque}, Rafael and {Basant}, Ritvik and {Brown}, Nina and {Das}, Tanya and {Kasper}, David and {Piaulet-Ghorayeb}, Caroline and {Kempton}, Eliza M.-R. and {Kite}, Edwin},
        title = "{The JWST Rocky Worlds DDT Program Reveals GJ 3929b to Likely Be a Bare Rock}",
      journal = {\apjl},
         year = 2025,
        month = dec,
       volume = {995},
       number = {2},
          eid = {L52},
        pages = {L52},
          doi = {10.3847/2041-8213/ae2098},
archivePrefix = {arXiv},
       eprint = {2508.12516},
 primaryClass = {astro-ph.EP},
       adsurl = {https://ui.adsabs.harvard.edu/abs/2025ApJ...995L..52X}
}

@ARTICLE{MeierVald2025,
       author = {{Meier Vald{\'e}s}, E.~A. and {Demory}, B.-O. and {Diamond-Lowe}, H. and {Mendon{\c{c}}a}, J.~M. and {August}, P.~C. and {Fortune}, M. and {Allen}, N.~H. and {Kitzmann}, D. and {Gressier}, A. and {Hooton}, M. and {Jones}, K.~D. and {Buchhave}, L.~A. and {Espinoza}, N. and {Fisher}, C.~E. and {Gibson}, N.~P. and {Heng}, K. and {Hoeijmakers}, J. and {Prinoth}, B. and {Rathcke}, A.~D. and {Eastman}, J.~D.},
        title = "{Hot Rocks Survey: II. The thermal emission of TOI-1468 b reveals a bare hot rock}",
      journal = {\aap},
         year = 2025,
        month = jun,
       volume = {698},
          eid = {A68},
        pages = {A68},
          doi = {10.1051/0004-6361/202453449},
archivePrefix = {arXiv},
       eprint = {2503.19772},
 primaryClass = {astro-ph.EP},
       adsurl = {https://ui.adsabs.harvard.edu/abs/2025A&A...698A..68M}
}

@ARTICLE{Patel2024,
       author = {{Patel}, J.~A. and {Brandeker}, A. and {Kitzmann}, D. and {Petit dit de la Roche}, D.~J.~M. and {Bello-Arufe}, A. and {Heng}, K. and {Meier Vald{\'e}s}, E. and {Persson}, C.~M. and {Zhang}, M. and {Demory}, B.-O. and {Bourrier}, V. and {Deline}, A. and {Ehrenreich}, D. and {Fridlund}, M. and {Hu}, R. and {Lendl}, M. and {Oza}, A.~V. and {Alibert}, Y. and {Hooton}, M.~J.},
        title = "{JWST reveals the rapid and strong day-side variability of 55 Cancri e}",
      journal = {\aap},
         year = 2024,
        month = oct,
       volume = {690},
          eid = {A159},
        pages = {A159},
          doi = {10.1051/0004-6361/202450748},
archivePrefix = {arXiv},
       eprint = {2407.12898},
 primaryClass = {astro-ph.EP},
       adsurl = {https://ui.adsabs.harvard.edu/abs/2024A&A...690A.159P}
}
\bibliographystyle{aasjournalv7}
\end{document}